\pdfoutput=1

\documentclass[
  twocolumn,
  prb,
  showpacs,
  amsmath,
  amssymb,
  superscriptaddress
]{revtex4}

\usepackage{bm}
\usepackage{graphicx}
\usepackage{color}
\usepackage{braket}
\usepackage{ulem}
\usepackage{pifont}
\usepackage{scrextend}
\usepackage{MnSymbol}

\newcommand{\bkt}{\mathrm{BKT}}
\newcommand{\hn}{\mathrm{HN}}
\newcommand{\bcs}{\mathrm{BCS}}
\newcommand{\sit}{\mathrm{SIT}}

\newcommand{\tr}{\text{tr}}
\newcommand{\Tr}{\text{Tr}}

\renewcommand{\v}[1]{\textbf{\textit #1}}

\definecolor{lightgray}{gray}{.9}

\begin{document}

\title{Berezinskii-Kosterlitz-Thouless transition in homogeneously disordered superconducting films}

\author{E.\ J.\ K\"onig}
\affiliation{Institut f\"ur Nanotechnologie, Karlsruhe Institute of Technology, 76021 Karlsruhe, Germany}
\affiliation{Inst. f\"ur Theorie der kondensierten Materie, Karlsruhe Institute of Technology, 76128 Karlsruhe, Germany}
\affiliation{Department of Physics, University of Wisconsin-Madison, Madison, Wisconsin 53706, USA}

\author{A. Levchenko}
\affiliation{Department of Physics, University of Wisconsin-Madison, Madison, Wisconsin 53706, USA}\affiliation{Institut f\"ur Nanotechnologie, Karlsruhe Institute of Technology, 76021 Karlsruhe, Germany}

\author{I.\ V.\ Protopopov}
\affiliation{Institut f\"ur Nanotechnologie, Karlsruhe Institute of Technology, 76021 Karlsruhe, Germany}\affiliation{Inst. f\"ur Theorie der kondensierten Materie, Karlsruhe Institute of Technology, 76128 Karlsruhe, Germany}
\affiliation{L.\ D.\ Landau Institute for Theoretical Physics RAS, 119334 Moscow, Russia}

\author{I.\ V.\ Gornyi}
\affiliation{Institut f\"ur Nanotechnologie, Karlsruhe Institute of Technology, 76021 Karlsruhe, Germany}
 \affiliation{A.\ F.\ Ioffe Physico-Technical Institute, 194021 St. Petersburg, Russia}
 \affiliation{L.\ D.\ Landau Institute for Theoretical Physics RAS, 119334 Moscow, Russia}

\author{I.\ S.\ Burmistrov}
\affiliation{L.\ D.\ Landau Institute for Theoretical Physics RAS, 119334 Moscow, Russia}
\affiliation{Moscow Institute of Physics and Technology, 141700 Moscow, Russia}

\author{A.\ D.\ Mirlin}
\affiliation{Institut f\"ur Nanotechnologie, Karlsruhe Institute of Technology, 76021 Karlsruhe, Germany}
\affiliation{Inst. f\"ur Theorie der kondensierten Materie, Karlsruhe Institute of Technology, 76128 Karlsruhe, Germany}
\affiliation{Petersburg Nuclear Physics Institute, 188300 St.~Petersburg, Russia.}
\affiliation{L.\ D.\ Landau Institute for Theoretical Physics RAS, 119334 Moscow, Russia}

\begin{abstract}
We develop a theory for the vortex unbinding transition in homogeneously disordered superconducting films. This theory incorporates the effects of quantum, mesoscopic and thermal fluctuations stemming from length scales ranging from the superconducting coherence length down to the Fermi wavelength. In particular, we extend the renormalization group treatment of the diffusive nonlinear sigma model to the superconducting side of the transition. Furthermore, we explore the mesoscopic fluctuations of parameters in the Ginzburg-Landau functional. Using the developed theory, we determine the dependence of essential observables (including the vortex unbinding temperature, the superconducting density, as well as the temperature-dependent resistivity and thermal conductivity) on microscopic characteristics such as the disorder-induced scattering rate and bare interaction couplings.
\end{abstract}

\date{\today}


\maketitle

\section{Introduction}

Disordered superconductors, the superconductor-insulator quantum phase transition, and quantum transport through superconducting nanodevices remain a field of intense research over the past 50 years. On the one hand, superconducting electronic components are of great technological interest in connection with a variety of possible applications. On the other hand, the interplay of two most outstanding manifestations of quantum coherence---superconductivity and Anderson localization---determines the attention attracted by these systems in the context of fundamental physics research, see, in particular, Refs.~\onlinecite{GantmakherDolgopolov2010,Feigelman-Ioffe-Mezard,BurmistrovGornyiMirlin2015} and references therein.

Superconductivity, i.e., the phenomenon of frictionless transport and perfect diamagnetism, is a consequence of long-ranged correlations of the complex order parameter $\Delta(\v x)$ in a theory of charged particles:
\begin{equation}
\left \langle \Delta^*(\v x) \Delta(0) \right \rangle \stackrel{x \rightarrow \infty}{\sim} \left \lbrace 
\begin{array}{cc}
e^{-x/\xi_\Delta}, & \text{normal state;} \\ 
\vert \langle\Delta\rangle \vert^2,  & \text{superconductor}.
\end{array} \right .
\end{equation}
Here, $\xi_\Delta$ is the correlation length. In two spatial dimensions at finite temperature true long-range order is not possible in view of the Mermin-Wagner theorem. In this case, one resorts to the following weaker definition:
\begin{equation}
\left \langle \Delta^*(\v x) \Delta(0) \right \rangle \stackrel{x \rightarrow \infty}{\sim} \left \lbrace 
\begin{array}{cc}
e^{-x/\xi_\Delta}, & \text{normal state;} \\ 
1/x^\eta,   &\text{superconductor}.
\end{array} \right .
\end{equation}	
In the last equation,  the power $\eta$ takes values $0<\eta  < 1$.

Typically, the following two sufficient conditions are fulfilled in a superconductor: (i) The modulus of the expectation value $\vert \langle \Delta(\v x) \rangle\vert $ is nonvanishing and nearly homogeneous. (ii) Strong phase fluctuations of $\phi = \arg\left (\langle \Delta (\v x)\rangle \right )$ are suppressed due to sufficiently large phase rigidity.

As a consequence of these conditions, two different mechanisms driving the transition between the superconducting and the normal state are often distinguished: (i) the expectation value $ \langle \Delta(\v x) \rangle$ vanishes across the transition; (ii) the expectation value $ \langle \Delta(\v x) \rangle \neq 0$ is locally finite, but the phase rigidity vanishes across the transition.

The first of these two mechanisms is sometimes referred to as ``fermionic'' scenario. It includes the Bardeen-Cooper-Schrieffer (BCS) theory~\cite{BardeenCooperSchrieffer1957} and related theories. In contrast, in the second ``bosonic'' mechanism, the phase fluctuations of preformed Cooper pairs drive the transition; typically the fermionic spectrum displays a pseudogap even in the normal state. 			
					
A particularly important representative of bosonic theories is the Berezinskii-Kosterlitz-Thouless (BKT) transition~\cite{Berezinskii, KosterlitzThouless73} occuring in two dimensional (2D) films. In a system with broken $\mathbf U(1)$-symmetry the phase degree of freedom corresponds to a Goldstone boson, the latter being described by the following $\mathbf U(1)$ nonlinear sigma model (NL$\sigma$M) action:
\begin{equation}
S_{\mathbf U(1)} = \frac{K}{\pi} \int_{\v x} \left (\nabla \phi \right )^2. \label{eq:intro:O2model}
\end{equation}
Here and below we use a short-hand notation $\int_{\v x}$ for the spatial integral over the 2D system $\int d^2x$.
In a clean superconductor the phase rigidity is given by ${K}/{\pi } = {n_s}/{2m^*T}$, where the density of Cooper pairs of mass $m^* = 2m$ is denoted by $n_s$. We set both the Boltzmann and Planck constants to unity, $k_B = \hbar =1$, throughout the paper. 

Since the manifold of Goldstone bosons is flat, the $\mathbf U(1)$ NL$\sigma$M, Eq.~\eqref{eq:intro:O2model}, is not renormalized perturbatively. In other words, the theory is Gaussian on the perturbative level. However, the inclusion of nonperturbative effects (vortices) leads to a two-parameter renormalization group (RG).\cite{JoseNelson1977} If the stiffness is less than the critical value $K_* = 1$, vortices proliferate and the system can not sustain phase coherence. Contrary, for $K> K_*$ there is a regime where vortices are tightly bound into dipoles and the system is superconducting. The critical stiffness defines the BKT transition temperature $T_\mathrm{BKT}$. If we assume $n_s$ to be determined by the BCS expression, $T_\bkt$ is parametrically close to the BCS transition temperature $T_{\rm BCS}$:
\begin{equation}
T_\bkt= T_{\rm BCS} \left (1 - 4 \, Gi \right ). \label{eq:intro:Tstar}
\end{equation}
Here, $Gi = T_{\rm BCS}/\mu \ll 1$ is the Ginzburg-Levanyuk number~\cite{Footnote-Gi} for clean 2D superconductors\cite{LarkinVarlamov2005} and $\mu$ denotes the chemical potential.		
					
In this paper we concentrate on disordered systems, where the elastic scattering rate $1/\tau$ of the electrons satisfies the condition 
\begin{equation}
T_\bkt \ll 1/\tau \ll \mu. \label{eq:intro:DefDiffusiveRegime}
\end{equation} 
According to ``Anderson's theorem'',\cite{AbrikosovGorkov1958,Anderson1959} the critical temperature $T_{c}$ with and without disorder are equal $T_c = T_{\rm BCS}$ in the absence of electron-electron interaction in singlet or triplet channels. The ``Anderson theorem'' discards, however, two important quantum-interference phenomena that can dramatically affect the superconductivity. The first one is the disorder-induced Anderson localization.\cite{Anderson1958} The localization is a natural antagonist of superconductivity, and a competition between them leads to emergence of superconductor-insulator transition in 2D systems. 
Second, the interelectronic Coulomb interaction can drastically change the critical temperature $T_c$, see Ref.~\onlinecite{Ovchinnikov1973} for an early work in this direction. The analysis of RG equations for the interacting NL$\sigma$M of diffusive soft modes\cite{Finkelstein1983,Finkelstein1984, Finkelstein1990, BelitzKirkpatrick1994, Finkelstein2010} predicts suppression of $T_c$, which is governed by the following expression\cite{MaekawaFukuyama1982,TakagiKuroda1982,Finkelstein1987,FinkelsteinSITreview}
\begin{equation}
\frac{T_c}{T_{\bcs}} =\exp\left(-\frac{1}{\gamma_c^{(0)}}\right) \left (\frac{\gamma_c^{(0)}+\sqrt{t_D}/2}{\gamma_c^{(0)}- \sqrt{t_D}/2}\right )^{\frac{1}{\sqrt{t_D}}} . \label{eq:intro:FinkelsteinSuppresion}
\end{equation}		
In this formula $\gamma_c^{(0)} = \frac{1}{\ln(T_{\bcs}\tau)}$. The result is valid in the limit of sufficiently large dimensionless Drude conductance $g_D = 2/(\pi t_D) \gg 1$ in units of $e^2/h$. Finally, for short range interaction, the mean-field transition temperature is predicted to increase in the presence of disorder,\cite{BurmistrovGornyiMirlin2012,DellAnna2013, BurmistrovGornyiMirlin2015,MayohGarciaGarcia2015} (provided $\vert \gamma_c^{(0)}\vert  \ll t$) 
\begin{equation}
\frac{T_c}{T_{\bcs}} \simeq \exp\left(-\frac{1}{\gamma_c^{(0)}}- \frac{2}{t_D}\right).\label{eq:intro:EnhancementTC} 
\end{equation}
The physical mechanism behind this phenomenon is wave function multifractality, which leads to an enhancement of the matrix elements of interaction.\cite{FeigelmanYuzbashyan2007,FeigelmanCuevas2010}

The goal of this paper is to develop a theory of the BKT transition in disordered superconductors that takes into account the localization and the Coulomb-interaction physics. Let us first remind the reader about key previous works that, while having strongly advanced the understanding of the disordered BKT transitions, did not include effects associated with the localization and the Coulomb interaction.
		
In full analogy to the clean case, the disordered BKT transition is determined by $K_* = 1$, with the only difference being
\begin{equation}
K= \frac{\pi g_D}{16} \frac{\Delta}{T} 
\tanh \frac{\Delta}{2T} \label{eq:intro:StiffnessBare}
\end{equation}
in the case of disordered films. As a consequence, in the disordered case,\cite{BeasleyMooijOrlando1979} $T_\bkt$ is again given by Eq.~\eqref{eq:intro:Tstar} with $Gi$ now being\cite{LarkinVarlamov2005} 
\begin{equation}
Gi = \frac{7 \zeta(3) }{\pi^3 g_D} \ll 1.\label{eq:intro:BareGiNumber}
\end{equation}
												
Most prominently, the superconducting transition manifests itself in the temperature dependence of conductance. There is a parametrically small temperature window close to $T_\bkt$ within which the normal state resistance rapidly drops to exponentially small values.  Above $T_{c}$, fluctuating Cooper pairs lead to an enhancement of conductivity, via Aslamazov-Larkin,\cite{AslamazovLarkin1968} Maki-Thompson\cite{Maki1968,Thompson1970} and density of states\cite{AltshulerVarlamovReizer1983} (DOS) corrections.\cite{LarkinVarlamov2005}
In the superconducting state, the resistance is determined by vortex excitations in the order parameter field.\cite{HalperinRefaelDemler2011} Due to the Josephson relation $\dot \phi = 2 e V$ a finite resistance (steady voltage $V$) requires a phase relaxation mechanism. In the 2D case close to the thermodynamic transition, the phase relaxation rate is determined by the vortices traveling across the system perpendicularly to the current direction.\cite{AmbegaokarSiggia1978,AmbegaokarSiggia1980}  Above the vortex unbinding temperature, the latter proliferate and the following crossover formula for the resistance $\rho$ near $T_c$ was suggested by Halperin and Nelson in Ref.~\onlinecite{HalperinNelson1979} 
\begin{equation}
{\rho(T)} =  t_D \frac{1}{1 +  A_v\left (\frac{\xi_{\hn}(T)}{\xi (T_\bkt)}\right )^2}. 
\label{eq:NelsonHalperin}
\end{equation}
The numerical constant $A_v$ is fixed to $A_v = \pi^4/56 \zeta(3) \approx 1.45$, so that Eq.\eqref{eq:NelsonHalperin} interpolates between the vortex-generated resistance in the superconducting state and the Aslamazov-Larkin contribution above $T_c$. 
Here the length $\xi_{\hn} (T)$  is given by
\begin{subequations}
\begin{equation}
\xi_{\hn} (T) = \frac{\xi(T_\bkt)}{b} \sinh \left (b\sqrt{\frac{Gi}{\tau_\bkt}}\right ) \label{eq:introHNCorrellength}
\end{equation}
and provides an interpolation between the BKT coherence length 
\begin{equation}
\xi_\hn \to \xi_\bkt(T)=\frac{\xi(T_\bkt)}{2b}\exp(b\sqrt{Gi/\tau_\bkt})  \label{eq:IntroBKTCorrellength}
\end{equation}
in the limit $\tau_\bkt\ll Gi$ and the mean field behavior 
\begin{equation}
\xi_\hn\to\xi_\mathrm{GL}(T)=\frac{\xi(0)}{\sqrt{\tau_\bkt}}, \label{eq:IntroGLCorrellength}
\end{equation}
\label{eq:IntroCorrellength}
\end{subequations}
in the opposite limit, $\tau_\bkt\gg Gi$. The quantity $b$ is a fitting parameter of order unity.\cite{BenfattoCastellaniGiamarchi2009}
Further, $\xi (T_\bkt)$ is the Ginzburg-Landau coherence length given by $\xi(T_\bkt)=\xi(0)/\sqrt{Gi}$, and $\tau_\bkt$ denotes a relative distance to the BKT transition,
\begin{equation}
\tau_\bkt=(T-T_\bkt)/T_\bkt. 
\end{equation}

The observation of the BKT transition in superconducting films is a longstanding experimental challenge. The major difficulties are due to (a) the parametrically small window $T_c - T_\bkt \ll T_c$ and (b) finite size effects. Both the finite thickness\cite{Pearl1964} and the finite area\cite{Kogan2007,BenfattoCastellaniGiamarchi2009}  of the film lead to an infrared cut-off in the logarithmic vortex-antivortex interaction and may thus obscure the BKT physics. Defining features for the experimental observation of the BKT transition  are (i) a universal crossover function $\ln(\rho(T)/t_D) \propto 1/\sqrt{\tau_\bkt}$, cf. Eq.~\eqref{eq:NelsonHalperin}; (ii) a jump from linear to nonlinear resistance ($V \propto I^\alpha$) with $\alpha = 3$ right at $T_\bkt$; (iii) a crossover from sublinear to superlinear magnetoresistance. 
For early work in this area, we refer the reader to Refs.~\onlinecite{BeasleyMooijOrlando1979,Fiory83,Hebard83,Hsu92} and to the review \onlinecite{Minnhagen1987}. 
Recent years have witnessed a strong rise of experimental interest to this phenomenon, with observations of the BKT transition in superconducting films made of various materials, including indium oxide,\cite{Crane07,Liu11,Misra13} MoGe,\cite{Misra13}
titanium nitride,\cite{BaturinaVinokur2012}
niobium selenide,\cite{El-BanaBending2013} lead,\cite{ZhaoXue2013} niobium nitride,\cite{YongSiegel2013,KoushikGhosh2013,GangulyBenfatto2015} and iron selenide.\cite{Schneider2014} 

It is important to emphasize that the experimentally determined $T_\bkt$ may differ very substantially from Eqs.~\eqref{eq:intro:Tstar} with an input from Eq.~\eqref{eq:intro:BareGiNumber}; see, in particular, Ref.~\onlinecite{BaturinaVinokur2012}. 
The reasons for the insufficiency of the theory leading to
 Eqs.~\eqref{eq:intro:Tstar} and \eqref{eq:intro:BareGiNumber} is closely related to those of the failure of Anderson's theorem, see a discussion and references above. These equations are based on the expression \eqref{eq:intro:StiffnessBare} for the superconducting density that does neither take into account the strong renormalization of conductivity due to interference and interaction effects, nor the renormalization of $T_c$, see, in particular, Refs.~\onlinecite{HebardKotliar1989,SmithAmbegaokar1992}. As a consequence, Eqs.~\eqref{eq:intro:Tstar} and \eqref{eq:intro:BareGiNumber} are not sufficient to correctly predict $T_\bkt$ as a function of microscopic parameters encoded in $\tau \propto g_{D}$ and $T_{\rm BCS}$.  

We are thus facing the following questions: (i) What is the superconducting stiffness including the disorder- and interaction-induced corrections? (ii) What is the vortex unbinding temperature for homogeneously disordered superconducting films? (iii) What is the temperature dependence of resistivity? The goal of the present paper is to develop a theory that answers these questions. As we will show, this requires an implementation of a strategy that allows one to go from the Fermi-liquid theory at relatively high energies to the low-energy $\mathbf{U}(1)$ theory through a sequence of intermediate-scale effective field theories. We will assume throughout the paper, that the dimensionless resistance is small for all temperatures, $\rho(T) \ll 1$. 

The paper is structured as follows. Section~\ref{chap:determinationTBKT} is devoted to the determination of the vortex unbinding temperature $T_\bkt$. As will be explained in detail in Sec.~\ref{sec:Hierarchy}, our formalism is based on the consecutive use of Fermi Liquid (Sec.~\ref{sec:FLtheory}), diffusive NL$\sigma$M (Sec.~\ref{sec:NLSMRG}), Ginzburg-Landau (Sec.~\ref{sec:GLtheory}) and $\mathbf U(1)$ NL$\sigma$M (Sec.~\ref{sec:O2model}) theories. The subsequent Section \ref{sec:resistivity} is devoted to the temperature dependence of resistivity of the metallic film close to the superconducting transition. We conclude the paper with a summary and outlook. The most technical details of our calculations are delegated to a number of appendices.

\section{Field theory of disordered superconductors}\label{chap:determinationTBKT}

This section is devoted to the theoretical framework of the present paper and of disordered superconductors in general. It is instructive first to get a feeling for the energy and respective length scales in the problem (Sec.~\ref{sec:Hierarchy}). From the hierarchy of length scales, the Fermi liquid (Sec.~\ref{sec:FLtheory}), the diffusive NL$\sigma$M (Sec.~\ref{sec:NLSMRG}), the Ginzburg-Landau theory (Sec.~\ref{sec:GLtheory}) and the $\mathbf U(1)$ NL$\sigma$M (Sec.~\ref{sec:O2model}) appear as a sequence of theories. This section is structured following this hierarchy and in each subsection \ref{sec:FLtheory}-\ref{sec:O2model} we discuss the derivation and the perturbative renormalization of the corresponding theory.

The physical observables discussed in this section are the vortex-unbinding temperature $T_\bkt$ in a disordered superconductor as well as the superconducting density, see Sec.~\ref{sec:O2model}.

\subsection{Strategy and hierarchy of length scales}
\label{sec:Hierarchy}

We will be interested in temperatures close to the vortex unbinding transition: $\vert  \tau_\bkt \vert\ll 1$. Our calculations are controlled in the limit, when the normal state conductance close to the transition is large: $g \gg 1$.  In this case, $T_\bkt$ turns out to be parametrically close to the mean field transition temperature $T_{\rm MF}$. As we will discuss below [see specifically Sec.~\ref{sec:Tstar:GL:PertTheory} and Eq.~(\ref{eq:GL:TMF})], the latter is close but remains below the critical temperature $T_c$ associated with the BCS-like instability. Note that in general, $g \neq g_D$ and $T_c \neq T_{\rm BCS}$, see, e.g.,  Eqs.~\eqref{eq:intro:FinkelsteinSuppresion} and \eqref{eq:intro:EnhancementTC}. 

On length scales larger than the elastic mean free path $l = v_F \tau$ ($v_F$ is the Fermi velocity) we will self-consistently associate a length scale $L_E$  to an energy $E$ by $L_E = \sqrt{D(L_E)/E}$ and reversely define $E_L = D(L)/L^2$. Here, $D(L_E)$ is the diffusion constant at the scale $L_E$ and $D(l)=v_F l/2$.  Thus, the three length scales $L_T \sim L_{T_\bkt} \sim L_{T_{\rm MF}}$ are close to each other in the regime under consideration. This regime further implies a hierarchy of length scales that we shall expose in this section. This hierarchy is associated with a step-by-step quantum to classical crossover governed by subsequent freezing of excitations. 

Fermionic Landau quasiparticles with well-defined momentum are good excitations only on length scales shorter than the mean free path $l$, see Sec.~\ref{sec:FLtheory}. On longer length scales they are ``confined'' in diffusive soft modes: noninteracting diffusons and cooperons as well as interacting bosonic modes in the Cooper and particle-hole interaction channels. In general, their interplay leads to strong renormalization of the conductivity and of the mean field transition temperature $T_{\rm MF}$. In our approach, this effect will be captured by the RG technique applied to the interacting, diffusive NL$\sigma$M. The RG stops at a length scale $L_{T_c}$ parametrically close to $L_{T_{\rm MF}}$  (more detailed explaination can be found in  Sec.~\ref{sec:NLSMRG}). The only modes that remain at larger distances are static fluctuations of the order parameter field $\Delta( \v x)$. Thus, at the scale $L_{T_{c}}$ we derive the Ginzburg-Landau (GL) free energy functional weighting these fluctuations, see Sec.~\ref{sec:GLtheory}. Our derivation, which contains terms beyond leading order in $1/g$, also yields random fluctuations of the coefficents in the GL functional. For $T<T_{\rm MF}$ the Higgs field [fluctuations in the modulus of $\Delta (\v x)$] is gapped on the scale of the GL coherence length $\xi =\xi_{\rm GL} (T) \gg L_{T_{\rm MF}}$. In the symmetry broken state, only phase fluctuations of the order parameter field are important on scales exceeding $\xi$. Thus, at the scale of the coherence length we derive the disordered model of phase fluctuations, Sec.~\ref{sec:O2model}  (see also Ref.~[\onlinecite{Kravtsov-Oppermann}] for a related study). Disorder terms turn out to be RG-irrelevant and can be taken into account in a perturbative manner. This eventually leads to an action analogous to Eq.~\eqref{eq:intro:O2model}, but with a renormalized stiffness $K$. For $T_\bkt<T<T_{\rm MF}$, the theory also predicts a renormalized coherence length $\xi_{\rm BKT} > \xi$ beyond which phase correlations decay exponentially.

\begin{figure*}
\includegraphics[scale=.65]{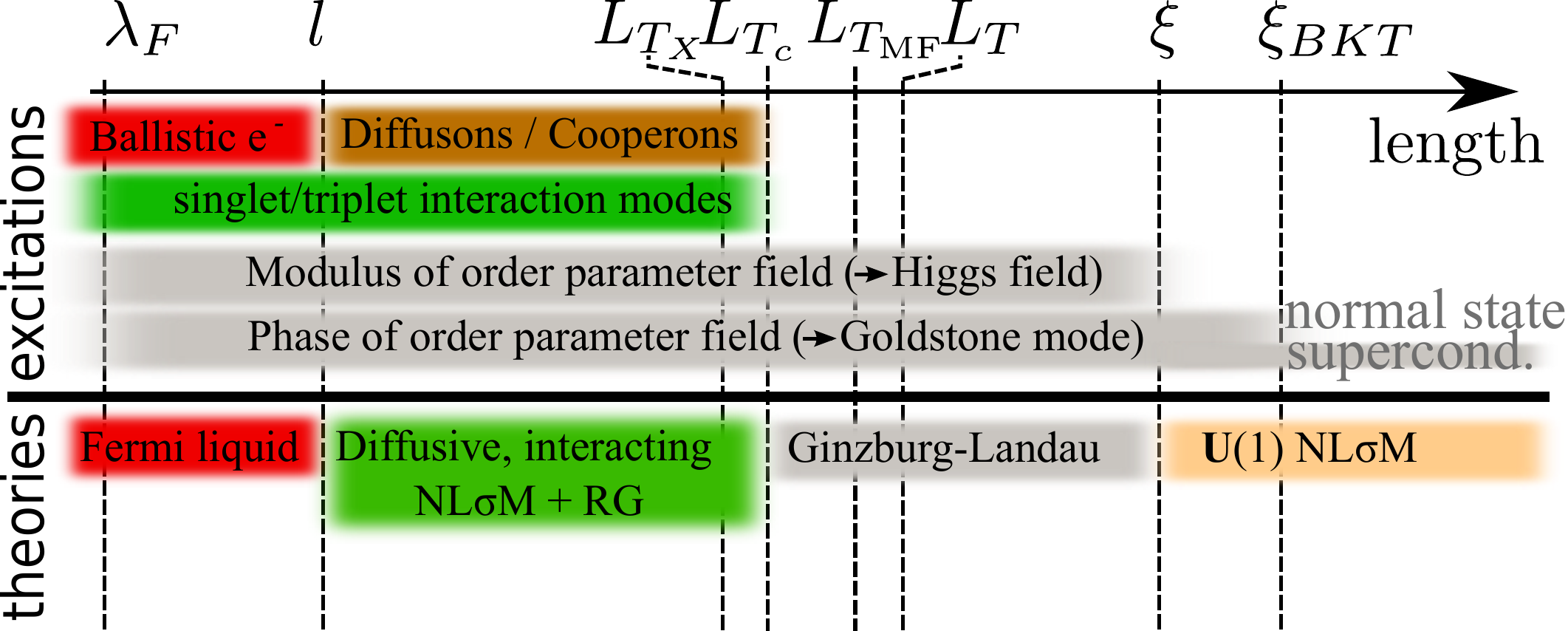}
\caption{Hierarchy of length scales for $T \simeq T_\bkt < T_{\rm MF}$. The scales $L_{T_X}$ and $L_{T_c}$ are introduced for technical reasons, see Sec.~\ref{sec:NLSMRG}, below. Both of them and $L_{T_\bkt}$ are parametrically close to $L_{T_{\rm MF}}$. In the category `excitations', a colored bar indicates the regime of importance of the various excitation modes. Analogously, in the category `theories', a colored bar indicates the regime, where a certain theory provides an appropriate description. A general explanation of the hierarchy of length scales can be found in the main text, Sec.~\ref{sec:Hierarchy}. }
\label{fig:lengthscales}
\end{figure*}

All in all, we find the following hierarchy of length scales (the Fermi wavelength is denoted by $\lambda_F$): 
\begin{equation}
\lambda_F < l < L_{T} < \xi < \xi_{\bkt}. \label{eq:lengthscales}
\end{equation}
At each intermediate length scale, a certain ``microscopic" theory ceases to be the appropriate description and we derive an emergent effective theory valid at longer length scales:
\begin{center}
\begin{tabular}{r|r c l}
scale & ``microtheory'' & & emergent theory \\
\hline $l$ & Fermi liquid &$\rightarrow$ & diffusive NL$\sigma$M, \\
 $L_{T_{c}}$ & diffusive NL$\sigma$M & $\rightarrow$ & GL theory, \\
 $\xi$ & GL theory & $\rightarrow$ & $\mathbf U(1)$ NL$\sigma$M.
\end{tabular}
\end{center}
A summary of length scales, relevant excitations and effective theories applicable to the various regimes is given in Fig.~\ref{fig:lengthscales}.

\subsection{Fermi liquid theory}
\label{sec:FLtheory}

The main statement of Landau's Fermi liquid theory is that, in the absence of spontaneous symmetry breaking, the low-energy excitations of a strongly correlated fermionic system are fermions (Landau quasiparticles) with the same quantum numbers as the free particles. Their decay rate is small as compared to the Fermi energy. In field-theoretical language, this statement means that the exact electronic Green's function (i.e. two point correlator) can be shown to contain a singular part (quasiparticle pole) with a weight $0<a<1$ and an additional regular contribution. \cite{Luttinger1961} 
			
Another particularly important quantity in the theory of strongly interacting fermions are the four point correlators.\cite{Nozieres_Luttinger_1962, AbrikosovGorkovDzyaloshinski1963, LandauLifshitz9} These implicitly define the full interaction amplitudes. The latter are subdivided into different channels of small energy-momentum transfer according to their tensor structure in spin space. For the problem of disordered, interacting, spinful fermions we concentrate on the static part of the Cooper singlet ($\Gamma_c$), particle-hole singlet ($\Gamma_s$) and particle-hole triplet ($\Gamma_t$) scattering amplitudes and keep only their zeroth angular harmonic ($s$-wave).\cite{Finkelstein1990, Finkelstein2010} In this paper, the quasiparticle residue $a$ is absorbed into the definition of fermionic fields and scattering amplitude.
			
Even though our goal is to describe a superconducting system close to and below $T_{\rm MF}$, i.e. in the symmetry broken phase, it is appropriate and justified to describe it using the Fermi liquid theory at the smallest length scales $L < l$. The fermionic excitations at these scales do not ``know'' about the fact that at larger length scales they will eventually form coherent Cooper pairs. In other words, if the system had a linear dimension $L_\square<l$, the associated size-quantization energy scale would by far exceed the superconducting gap.  
		
\subsection{The diffusive NL$\sigma$M}
\label{sec:NLSMRG}

In the previous section, we explained that the Fermi-liquid description is appropriate for disordered superconducting films at smallest length scales, i.e., those less than the elastic mean free path $l$. In our hierarchy of length scales we now reach the next level characterized by scales exceeding $l$. The effective field theory that emerges in this regime is the NL$\sigma$M of diffusive interacting soft modes (for review see Refs.~[\onlinecite{Finkelstein1990,BelitzKirkpatrick1994}]).

\subsubsection{Normal state NL$\sigma$M}
\label{sec:normal-nlsm}

Upon inclusion of sufficiently weak disorder (in the sense $g_D \gg 1$) static quantities, such as the static interaction amplitudes, remain unchanged even at scales $L \gg l$. They are determined by scales much shorter than the mean free path. On the contrary, the dynamical properties of the system are altered, as the retarded-advanced ladders consisting of two Green's functions acquire a diffusive pole.\cite{Finkelstein1990, Finkelstein2010} It is possible to describe the diffusive dynamics by means of the interacting, diffusive NL$\sigma$M.\cite{Finkelstein1983,Finkelstein1984} In the normal state, the path integral representation of the partition sum 
\begin{equation}
\mathcal Z = \int \mathcal D Q \exp(-S[Q])
\end{equation}
is governed by the following action:

\begin{subequations}
\begin{equation}
S = S_\sigma + S_{\rm int}^{(\rho)} + S_{\rm int}^{(\sigma)}+ S_{\rm int}^{(c)}, \label{eq:TStar:NLSM:FullNLSMBigStructure}
\end{equation}
with
\begin{eqnarray}
&&\hskip-.5cm
S_\sigma = \frac{g}{32} \int_{\v x} \tr \left [(\nabla Q)^2 \right ] - 2Z_\omega \int_{\v x} \tr [\hat \epsilon Q] , \label{eq:Tstar:NLSM:Ssigma}\\
&&\hskip-.5cm
S_{\rm int}^{(\rho)} = \frac{\pi T}{4} \Gamma_s \sum_{\substack{\alpha, n\\r = 0,3}} \int_{\v x} \tr\left [I_n^\alpha t_{r0} Q\right ] \tr\left [I_{-n}^\alpha t_{r0} Q\right ], \label{eq:Tstar:NLSM:Sintrho}\\
&&\hskip-.5cm
S_{\rm int}^{(\sigma)} = \frac{\pi T}{4} \Gamma_t \sum_{\substack{\alpha, n\\r = 0,3 \\ j = 1,2,3}}\int_{\v x} \tr\left [I_n^\alpha t_{rj} Q\right ] \tr\left [I_{-n}^\alpha t_{rj} Q\right ], \label{eq:Tstar:NLSM:Sintsigma}\\
&&\hskip-.5cm
S_{\rm int}^{(c)} = \frac{\pi T}{4} \Gamma_c \sum_{\substack{\alpha, n \\ r = 1,2}} \int_{\v x}\tr\left [L_n^\alpha t_{r0} Q\right ] \tr\left [L_{n}^\alpha t_{r0} Q\right ].\label{eq:Tstar:NLSM:SintCooper}
\end{eqnarray}\label{eq:Tstar:NLSM:FullNLSM}
\end{subequations}
Here, the symbol `tr' denotes summation over all internal matrix indices. 
We are interested in systems where time-reversal symmetry is fulfilled. If the system is additionally spin-rotational invariant (which corresponds to class AI in the classification of non-interacting systems), the $Q$ matrices are symplectic, traceless, and have nontrivial structure in replica, Matsubara, spin and Nambu spaces: 
\begin{subequations}
 \begin{equation}
 Q = Q^\dagger = Q^{-1} = t_{12} Q^T t_{12}, \quad \tr\, Q = 0.
 \label{eq:Q-TRS}
 \end{equation}
 In the absence of spin rotational symmetry (noninteracting class AII), the $Q$ matrices are orthogonal,\cite{Footnote-Q} traceless, have nontrivial structure in replica, Matsubara, and Nambu spaces, and are proportional to the identity matrix in spin space:
\begin{equation}
 Q = Q^\dagger = Q^{-1} = t_{10} Q^T t_{10} \propto \mathbf 1_\sigma, \quad \tr\, Q = 0.
 \label{eq:Q-no-TRS}
 \end{equation}
\end{subequations}
We use the convention $t_{rj} = \tau_r \otimes \sigma_j$ where $\tau_r = (\mathbf 1_\tau, \vec \tau)$ are the identity and the Pauli matrices in Nambu space, while $\sigma_j = (\mathbf 1_\sigma, \vec \sigma)$ are those in spin space. Here and throughout the paper we use a convention in which $\alpha, \beta =  1, \dots , N_R $ denote replicas and $m, n = - N'_M, \dots , N'_M - 1 $ Matsubara indices associated to fermionic frequencies $\epsilon_n = \pi T (2n + 1)$. The following matrices, which are trivial in Nambu and spin spaces, have been introduced:\cite{Footnote-Matsubara} 
\begin{subequations}
\begin{eqnarray}
\Lambda^{\alpha \beta}_{nm} &=& \text{sgn}\left (n \right )
 \delta^{\alpha \beta}  \delta_{nm} ,  \\
\hat \epsilon^{\alpha \beta}_{nm} &=& \epsilon_{n} \delta^{\alpha \beta} \delta_{nm}  , \\
\left (I^{\alpha_0}_{n_0} \right )^{\alpha \beta}_{nm} &=& \delta^{\alpha_0 \alpha} \delta^{\alpha_0 \beta}
\delta_{n-m,n_0}, \\
\left (L^{\alpha_0}_{n_0} \right )^{\alpha \beta}_{nm} &=& \delta^{\alpha_0 \alpha} \delta^{\alpha_0 \beta}
\delta_{n+m+1,n_0}. 
\end{eqnarray}			
 \label{eq:matrixdefs}
 \end{subequations}	
 
The coupling constants of the NL$\sigma$M are the dimensionless conductivity $g$ (bare value $g_D$), the static interaction amplitudes $\Gamma_i$ ($i = s,t,c$), and the prefactor $Z_\omega$ of the frequency term, which is related to the renormalization of specific heat. Note that the latter does not flow in the noninteracting case and keeps the bare value $Z_\omega^{(0)} = \pi \nu/4$. In the presence of long range Coulomb interaction the NL$\sigma$M is ``$\mathcal F$-invariant''. \cite{PruiskenBaranovSkoric1999} Essentially, this means electrostatic gauge invariance (i.e. invariance under time dependent but space independent phase rotations) and fixes $Z_\omega + \Gamma_s = 0$. In the present convention, attraction in the Cooper channel implies $\Gamma_c <0$.
				
For the sake of a better readability, we omit gauge potentials in Eq.~\eqref{eq:Tstar:NLSM:FullNLSM} and in the rest of the paper (except App.~\ref{app:Kappa}). Thus, we formally treat a neutral superfluid. As we explain in Sec.~\ref{sec:finitesize}, in the truly 2D limit all presented results hold equally for a charged superconductor.

\subsubsection{NL$\sigma$M in the superconducting state}
\label{sec:SCNLSM}

While in the normal state diffusive fluctuations are associated to smooth variations around the noninteracting saddle point solution $Q = \Lambda$, in the symmetry broken phase the true saddle point is a function of the superconducting gap. This result can be directly derived from the microscopic theory, see Refs.~\onlinecite{YurkevichLerner2001a,FLS,ALAK,Koenig2014}. However, in such a derivation the disorder is taken into account at the level of Anderson's theorem only. Here we derive the NL$\sigma$M for the symmetry broken phase directly from the interacting normal state NL$\sigma$M, Eqs.~\eqref{eq:Tstar:NLSM:FullNLSM}. Since this procedure can be accomplished at any scale, this allows us to go beyond Anderson's theorem.

Upon Hubbard-Stratonovich decoupling in the Cooper channel the action becomes 
\begin{eqnarray}
S_{\rm HS}[Q,\check \Delta] &=& \frac{g}{32} \Tr \left [(\nabla Q)^2 \right ] - 2Z_\omega \Tr Q\left [\hat \epsilon + i \left (\begin{array}{cc}
0 & - \check \Delta \\ 
\check \Delta ^\dagger & 0
\end{array} \right )\right ] \notag \\
&& - \frac{4Z_\omega}{\pi \gamma_c T} \sum_{\textcolor{black}{\alpha} \textcolor{black}{n}} \int_{\v x} \Delta^*_{\textcolor{black}{\alpha} \textcolor{black}{n}}\Delta_{\textcolor{black}{\alpha} \textcolor{black}{n}} + S_{\rm int}^{(\rho)} + S_{\rm int}^{(\sigma)}. \label{eq:NLSM:SSBNLSM:DecoupledNLSM}
\end{eqnarray}
The symbol `$\Tr$' includes both the trace operation in matrix space and spatial integration.
The complex Hubbard-Stratonovich field $ \check \Delta$ is defined by
\begin{equation}
\check \Delta = \sum_{\textcolor{black}{\alpha}, \textcolor{black}{m}}  \Delta_{\textcolor{black}{\alpha}, \textcolor{black}{m}}  L^\alpha_{ \textcolor{black}{m}}.
\end{equation} 
We also introduced the notation $\gamma_c = \Gamma_c/Z_\omega$. Since we expect static $s$-wave superconductivity, the Hubbard-Stratonovich field can be assumed to take the form $\Delta_{\alpha,n} = \Delta_{\alpha} \delta_{n,0}$ on mean field level. We will refer to $\Delta_{\alpha}$ as ``order parameter field".
	 
Variation of Eq.~\eqref{eq:NLSM:SSBNLSM:DecoupledNLSM} with respect to $\Delta^*_{\alpha,0}$ leads to the gap equation
\begin{equation}
\frac{\Delta_{\alpha}}{\gamma_cT} =- \frac{i \pi }{2} \tr\left [Q \frac{\tau_x - i \tau_y}{2} L^\alpha_{\textcolor{black}{0}} \right ] \label{eq:saddlepointFromNLSMBCS1}. 
\end{equation}
Using the constraint \eqref{eq:Q-TRS} [or \eqref{eq:Q-no-TRS}] on the $Q$-matrix, we find that the second term in
\eqref{eq:NLSM:SSBNLSM:DecoupledNLSM} modifies the saddle point $Q=\Lambda$ to $Q = \bar \Lambda$ with
\begin{equation}
\bar \Lambda =  \sum_{n\geq 0, \alpha} P^\alpha_{\vert n \vert} \left [\frac{\vert \epsilon_n \vert \Lambda_z}{\sqrt{\epsilon^2_n + \vert \Delta_\alpha \vert ^2}}+\frac{\left (\begin{array}{cc}
0 & -i \Delta_\alpha \\ 
i \Delta_\alpha^* & 0
\end{array} \right )_{(\tau)}\Lambda_x}{\sqrt{\epsilon^2_n + \vert \Delta_\alpha \vert ^2}}\right ].\label{eq:Tstar:Lambdabar}
\end{equation}
By $P_{\textcolor{black}{\vert n  \vert}}^{\textcolor{black}{\alpha}}$ we denote a projector on replica $\alpha$ and on a block with given modulus of the Matsubara frequency, matrices $\Lambda_{x,y,z}$ are Pauli matrices in this space.

On the saddle point level, it is possible to choose the order parameter field real and equal in all replicas
\begin{equation}
\Delta_{\textcolor{black}{\alpha}} = \Delta > 0, \;  \textcolor{black}{\alpha} = 1, \dots, N_R.
\end{equation} 
Then $\bar \Lambda$ has the following structure in Matsubara and Nambu spaces:
\begin{equation}
\bar \Lambda =  \left (\begin{array}{cccccc}
\ddots & 0 & 0 & 0 & 0 & \udots \\ 
0 & \frac{\epsilon_{\textcolor{black}{ 2 }} }{\sqrt{\epsilon_{\textcolor{black}{2}}^2 + \Delta^2}} & 0 & 0 & \frac{  \Delta\tau_y }{\sqrt{\epsilon_{\textcolor{black}{2}}^2 + \Delta^2}} & 0 \\ 
0 & 0 & \frac{\epsilon_{\textcolor{black}{ 1 }}}{\sqrt{\epsilon_{\textcolor{black}{1}}^2 + \Delta^2}} & \frac{\Delta\tau_y}{\sqrt{\epsilon_{\textcolor{black}{1}}^2 + \Delta^2}} & 0 & 0 \\ 
0 & 0 & \frac{\Delta\tau_y }{\sqrt{\epsilon_{\textcolor{black}{1}}^2 + \Delta^2}} & \frac{-\epsilon_{\textcolor{black}{ 1 }} }{\sqrt{\epsilon_{\textcolor{black}{1}}^2 + \Delta^2}} & 0 & 0 \\ 
0 & \frac{\Delta\tau_y}{\sqrt{\epsilon_{\textcolor{black}{2}}^2 + \Delta^2}} & 0 & 0 & \frac{- \epsilon_{\textcolor{black}{ 2 }} }{\sqrt{\epsilon_{\textcolor{black}{2}}^2 + \Delta^2}} & 0 \\ 
\udots & 0 & 0 & 0 & 0 & \ddots
\end{array} \right ).\label{eq:Tstar:Lambdabar1}
\end{equation}
In the limit $\Delta \rightarrow 0$ the usual diffusive form is restored: $\bar \Lambda \rightarrow \Lambda = \text{diag}(\mathbf 1,- \mathbf 1)$. 	

Returning to the generally complex $\Delta_\alpha$, it is possible to perform a Bogoliubov transformation,\cite{YurkevichLerner2001a} 
\begin{equation}
Q = U_\Delta^\dagger q U_\Delta, \label{eq:Tstar:NLSM:BogoliubovTrafo}
\end{equation}
such that the saddle point of the rotated field is again $q = \Lambda$. In the parametriztion $\Delta_\alpha= \vert \Delta_\alpha \vert e^{i\phi_\alpha}$ the unitary rotation matrix is
		\begin{equation}
		U_{\Delta} = \sum_{n \geq 0, \alpha} P_{\vert n \vert}^\alpha e^{i \frac{\phi_\alpha}{2} \tau_z} \left (\begin{array}{cc}
		\cos \psi_n^\alpha & \tau_y \sin \psi_n^\alpha \\ 
		-\tau_y \sin \psi_n^\alpha & \cos \psi_n^\alpha
		\end{array} \right )_{(\Lambda)} e^{-i \frac{\phi_\alpha}{2} \tau_z}.
		\end{equation}
We introduced the energy dependent rotation angle 
\begin{equation}
\cos\psi_n^\alpha = \frac{1}{\sqrt 2} \sqrt{1 + \frac{\vert \epsilon_n \vert}{\sqrt{\epsilon_n^2 + \vert \Delta_\alpha\vert^2}}}.
\end{equation}

After integration of fluctuations $ \delta \Delta_{\alpha,n} = \Delta_{\alpha,n} - \Delta_{\alpha} \delta_{n,0}$ around the mean field solution we arrive at the NL$\sigma$M describing the system at $T<T_{\rm MF}$. It has the standard structure exposed in Eq.~\eqref{eq:TStar:NLSM:FullNLSMBigStructure}. The terms $S^{(\rho)}_{\rm int} [ Q ]  =  S^{(\rho)}_{\rm int} [  U_\Delta^\dagger q U_\Delta ] $ and $S^{(\sigma)}_{\rm int} [ Q ]  =  S^{(\rho)}_{\rm int} [  U_\Delta^\dagger q U_\Delta ] $ are determined by Eqs.~\eqref{eq:Tstar:NLSM:Sintrho} and \eqref{eq:Tstar:NLSM:Sintsigma} respectively. The Cooper channel interaction term, Eq.~\eqref{eq:Tstar:NLSM:SintCooper} is slightly modified, as the static mean field solution is subtracted from $Q$:
\begin{equation}
S_{\rm int}^{(c)}  = \frac{\pi T}{4} \Gamma_c \sum_{\substack{\alpha, n \\ r = 1,2}} \int_{\v x}\tr\left [L_n^\alpha t_{r0} Q^{(n)}\right ] \tr\left [L_{n}^\alpha t_{r0} Q^{(n)}\right ].
\end{equation}
with $Q^{(n)} = Q - \bar \Lambda \delta_{n,0}$. The major modification concerns the dynamic part of the action, i.e. Eq.~\eqref{eq:Tstar:NLSM:Ssigma}, which is most conveniently written in the rotated basis,

\begin{equation}
S_\sigma [q] = \frac{g}{32} \Tr \left [(\nabla q )^2\right ] - 2Z_\omega \Tr[q \hat \varepsilon]. \label{eq:Tstar:SSBNLSM}
\end{equation}
We have introduced the matrix
\begin{equation}
\hat \varepsilon^{\alpha \beta}_{n,m} = \text{sign}(\epsilon_n) \varepsilon_n^\alpha \delta_{n,m} \delta^{\alpha, \beta} \text{ with } \varepsilon_n^\alpha =  \sqrt{\epsilon_n^2 + \vert \Delta_\alpha \vert ^2}.
\end{equation} 

\subsubsection{Saddle point equation}
\label{sec:Tstar:NLSM:SaddlepointEq}
	
Using Eq.~\eqref{eq:Tstar:Lambdabar1}, the gap equation \eqref{eq:saddlepointFromNLSMBCS1} becomes 
\begin{eqnarray}
\frac{\Delta}{T} &=& - \gamma_c \pi  \Delta \sum_{\textcolor{black}{n =-\infty}}^{\textcolor{black}{\infty}}  \frac{ 1}{\sqrt{\epsilon_{\textcolor{black}{n}}^2 + \Delta^2}} \notag \\
&\doteq & - \gamma_c \frac{\Delta}{ T}\int_{ \Delta }^\lambda d\epsilon \frac{\tanh \frac{\epsilon}{2T}}{\sqrt{\epsilon^2 - \Delta^2}}. \label{eq:superconductivity:FermionicGapEq}
\end{eqnarray}
This equation has the structure of the standard BCS equation. The symbol ``$\doteq$'' indicates equality of sum and integral upon appropriate ultraviolet (UV) regularization at the scale $\lambda$.	The solution of this equation determines a transition temperature
\begin{equation}
T_c = \lambda \exp(1/\gamma_c). \label{eq:TcDiffusivegeneral}
\end{equation}
All standard implications (e.g.~$\Delta (T)$, the DOS etc.) immediately follow analogously to the BCS case. However, it is important to keep in mind, that in general the Cooper channel interaction parameter $\gamma_c = \gamma_c(L)$ is strongly scale dependent and influenced by disorder and interactions in the particle-hole channels.

In the simplest approximation, the NL$\sigma$M is analyzed at bare level. The diffusive UV cut-off is $\lambda \sim 1/\tau$ and $\gamma_c$ is replaced by $\gamma_c (l)$. It already includes the ballistic renormalization from scales between the Debye wavelength $\lambda_D$ and the mean free path $l$. Within this simplified analysis, $T_c$ is determined by the energy scale of divergence of the solution 
\begin{equation}
\gamma_c(L) = \frac{1}{[\gamma_c(l)]^{-1} - \ln E_L\tau }
\end{equation}
of the BCS like\cite{Shankar1994} RG equation 
\begin{equation}
\frac{d \gamma_c}{dy} = -2 \gamma_c^2,  \label{eq:Tstar:BCSShankarRG}
\end{equation}
where
\begin{equation}
y = - \ln (E_L\tau)/2
\end{equation}
is the logarithm of the running RG scale. 
The transition temperature appearing in Eq.~\eqref{eq:TcDiffusivegeneral} simply becomes the BCS transition temperature $T_c = T_{\rm BCS}$ with
\begin{equation}
T_{c} = \frac{1}{\tau} e^{\frac{1}{\gamma_c(l)}} = \omega_D e^{\frac{1}{\gamma_c(\lambda_D)}}  = T_c e^{\frac{1}{\gamma_c(L_{T_c})}},\label{eq:BCStranstemp}
\end{equation}
where $\omega_D$ is the Debye frequency.
This equation is a restatement of Anderson's theorem. For BCS superconductors, $\gamma_c(\lambda_D)$ determines the bare phonon mediated interaction vertex. 
	
In general, disorder and interaction effects modify the RG equation \eqref{eq:Tstar:BCSShankarRG}. Then, the transition temperature $T_c$ (defined via the scale where $\gamma_c$ diverges) may strongly differ from $T_{\rm BCS}$.
	 
\subsubsection{Renormalization group flow}\label{sec:NLSM:RGdiscussion}

In the normal state, the noninteracting diffuson of a quasiparticle pair with Matsubara frequencies $\epsilon_{n_1},\epsilon_{n_2}$ ($n_1 \geq0$, $n_2<0$) and replica indices $\alpha, \beta$ is determined by the following diffusion propagator (see App.~\ref{App:GLDeriv})
			
\begin{equation}
\left [\mathcal D (\v q)\right ]_{n_1,n_2}^{\alpha \beta} = \frac{D}{D \v q^2 + \epsilon_{n_1} - \epsilon_{n_2}}. \label{eq:GL:NormalstatePropagator}
\end{equation}
The diffusion coefficient $D$ is determined by the coupling constants of the NL$\sigma$M via $D = g/16Z_\omega$. In the superconducting state this propagator becomes
\begin{equation}
\left [\mathcal D_\Delta(\v q)\right ]_{n_1,n_2}^{\alpha \beta} = \frac{D}{D \v q^2 + \varepsilon_{n_1}^\alpha + \varepsilon_{n_2}^\beta}. \label{eq:NLSM:InversePropzero}
\end{equation}
However, in the interval of length scales $l < L <  L_T \ll L_\Delta$ the effect of the superconducting gap on Eq.~\eqref{eq:NLSM:InversePropzero} is negligible. Therefore, in this interval, it is legitimate to construct a perturbative RG around the normal state saddle point $\Lambda$. (Fluctuations are too fast to resolve the difference between $\Lambda$ and $\bar \Lambda$.) Thus, in this regime of length scales, the renormalization of parameters in superconducting state is dictated by the RG equations of the normal state NL$\sigma$M, Eq.~\eqref{eq:Tstar:NLSM:FullNLSM}. These have the following form 
\begin{subequations}
\begin{eqnarray}
&&\frac{dt}{d y} =\beta_t\left (t, \gamma_s,\gamma_t,\gamma_c \right ), \\
&&\frac{d\gamma_s}{d y} = \beta_{\gamma_s}\left (t, \gamma_s,\gamma_t,\gamma_c\right ), \\
&&\frac{d\gamma_t}{d y} = \beta_{\gamma_t}\left (t, \gamma_s,\gamma_t,\gamma_c\right ), \\
&&\frac{d\gamma_c}{d y} = \beta_{\gamma_c}\left (t, \gamma_s,\gamma_t,\gamma_c\right ).
\end{eqnarray}
					
It is worth stressing that, as a consequence of dimensional analysis, the RG equations can be written in terms of reduced coupling constants $\gamma_i = \Gamma_i/Z_\omega$ (with $i = s,t,c$). The particle-number conservation implies that the combination $Z_\omega+\Gamma_s$ is not renormalized. Therefore 
\begin{equation}
\frac{d \ln (Z_\omega)}{dy} = - \frac{ \beta_{\gamma_s}\left (t, \gamma_s,\gamma_t,\gamma_c\right )}{1+\gamma_s}. \label{eq:Tstar:RGequationsAI:lnzdot}
\end{equation} \label{eq:Tstar:GeneralRGequations}
\end{subequations}

In Ref.~\onlinecite{Finkelstein1984}, the beta functions, Eqs.~\eqref{eq:Tstar:GeneralRGequations}, were derived to the lowest order in $t$ and $\gamma_c$. Recently,\cite{BurmistrovGornyiMirlin2015} three of us extended the results of Ref.~\onlinecite{Finkelstein1984} by deriving one-loop (lowest order with respect to $t$) beta functions which are formally exact in the Cooper channel interaction constant $\gamma_c$.  Here, we only quote the result and refer to Ref.~\onlinecite{BurmistrovGornyiMirlin2015} for more details:
\begin{subequations}
\begin{eqnarray}
&&\hskip-.65cm
\beta_{t}=t^2 \left [1 + f(\gamma_s) + 3 f(\gamma_t) - \gamma_c \right ], \label{eq:Tstar:RGequationsAI:tdot} \\
&&\hskip-.65cm
\beta_{\gamma_s} =- \frac{t}{2} \left (1+ \gamma_s\right )\left (\gamma_s + 3 \gamma_t + 2 \gamma_c + 4 \gamma_c^2\right ), \label{eq:Tstar:RGequationsAI:gammasdot}\\
&&\hskip-.65cm
\beta_{\gamma_t} = - \frac{t}{2} \left (1+ \gamma_t\right )\left [\gamma_s -\gamma_t  - 2 \gamma_c  \left (1+2\gamma_t-2\gamma_c\right )\right ], \label{eq:Tstar:RGequationsAI:gammatdot}\\
&&\hskip-.65cm
\beta_{\gamma_c} = - \frac{t}{2} \Big [\left (1+ \gamma_c\right )\left (\gamma_s - 3\gamma_t \right ) -2 \gamma_c^2 + 4 \gamma_c^3 \notag \\
&&\hskip-.65cm
+ 6 \gamma_c\left (\gamma_t-\ln \left (1+\gamma_t\right )\right )\Big ] - 2 \gamma_c^2. \label{eq:Tstar:RGequationsAI:gammacdot}
\end{eqnarray}\label{eq:Tstar:RGequationsAI}
\end{subequations}
The function $f(x)$ entering Eq.~\eqref{eq:Tstar:RGequationsAI:tdot} is given by
\begin{equation}
f(x) = 1 - \frac{1+x}{x} \ln(1+x).\label{eq:FundamentalsLocalization:Finkelstein:functionf}
\end{equation}
				
Equations \eqref{eq:Tstar:RGequationsAI} are appropriate for a system with spin-rotation invariance. The first term ``$1$'' in the square bracket of Eq.~\eqref{eq:Tstar:RGequationsAI:tdot} describes the weak-localization effect, which originates from disorder and is unrelated to interactions.  The last term ``$-2\gamma_c^2$'' in Eq.~\eqref{eq:Tstar:RGequationsAI:gammacdot} represents the Cooper instability, which is also present in clean systems.\cite{Shankar1994} All other terms stem from the interplay of disorder and interactions. 
				
In the case of a system with strong spin-orbit coupling the following modifications to Eqs.~\eqref{eq:Tstar:RGequationsAI} occur. First, one should replace the weak-localization by the weak-antilocalization effect, i.e. ``$1$'' in the square bracket of Eq.~\eqref{eq:Tstar:RGequationsAI:tdot} by ``$-1/2$''. Second, the triplet channel is gapped out, so that 
Eq.~\eqref{eq:Tstar:RGequationsAI:gammatdot} should be discarded and 
terms containing $\gamma_t$ should be removed from the remaining equations.

\subsubsection{Range of applicability of perturbative RG}	
\label{sec:RangeApplicability}
				
As we have already stated, per definition $\gamma_c(L_T) \stackrel{L_T\rightarrow L_{T_c}}{\longrightarrow } - \infty$. Thus, close to $T_c$, one may be tempted to keep only the leading powers of $\gamma_c$ in Eqs.~\eqref{eq:Tstar:RGequationsAI}. However, one should keep in mind that the RG Eqs.~\eqref{eq:Tstar:RGequationsAI} were derived in the one loop approximation (i.e. perturbatively in resistance $t$). An inspection of the perturbative series in the vicinity of $T_c$ (where $|\gamma_c|$ is large) indicates\cite{BurmistrovGornyiMirlin2015} that the actual parameter of the expansion in this region is $t|\gamma_c|$. 
Thus, close to $T_c$, the RG equations are only applicable for energy scales $E_L \gtrsim T_X$, where  $T_X$ is defined by 
\begin{equation}
\vert \gamma_c(L_{T_X}) t(L_{T_X}) \vert=1.
\label{eq:LX}
\end{equation} 
Therefore, disorder-induced corrections are subleading with respect to the dominant Cooper-instability term within the range of applicability of Eqs.~\eqref{eq:Tstar:RGequationsAI}.  Therefore, close to $T_c$, the Cooper channel coupling constant diverges as
\begin{equation}
\gamma_c (L_T) \stackrel{L_T\rightarrow L_{T_c}}{\sim} \frac{{T_c}}{{T_c}-T} \quad (L_T \lesssim L_{T_X} < L_{T_c}). \label{eq:Tstar:NLSM:BCSlike}
\end{equation}				
We remind the reader that $T_c$ (and thus $L_{T_c}$) as given by the RG equations~\eqref{eq:Tstar:RGequationsAI} is in general strongly renormalized as compared to $T_{\rm BCS}$, see Eqs.~\eqref{eq:intro:FinkelsteinSuppresion} and~\eqref{eq:intro:EnhancementTC} of the introduction.
				
We expect that the full RG equations~\eqref{eq:Tstar:GeneralRGequations} contain a line of attractive fixed points characterized by $\gamma_c(L_{T_c})= -\infty$ but finite $Z_\omega = Z_\omega(L_{T_c})$ and $t = t(L_{T_c}) \leq t_{\sit} \sim 1$. This expectation is based on the fact that $Z_\omega$ and $t$ determine the coefficients of the GL functional (see Sec.~\ref{sec:GLtheory} below). 
The line ends in the point with coupling constant $t_{\sit}$ which represents the quantum critical point of the superconductor-insulator transition. While it is worth emphasizing the difference between the physical resistance $\rho$ and the NL$\sigma$M coupling constant $t$ (see Sec.~\ref{sec:resistivity}  below and Ref. \onlinecite{BurmistrovGornyiMirlin2015}) at this quantum critical point we expect $\rho = t = t_{\sit}$.
				
We now present an estimate for the difference between ${T_c}$, $t(L_{T_c})$, $Z_\omega(L_{T_c})$ and $T_X$, $t(L_{T_X})$, $Z_\omega(L_{T_X})$, respectively.  We will assume that (a) there is a region of fixed points, (b) $\gamma_c^{-1} {\simeq}  \alpha_c \ln (T_c/T)$ as  $L_T \rightarrow L_{T_c}$ ($\alpha_c \sim 1$), (c) the crossover of RG flow from Eqs.~\eqref{eq:Tstar:RGequationsAI} to the fixed point region is encoded in a power series of the parameter $\vert \gamma_c t \vert$ entering the beta functions, Eq.~\eqref{eq:Tstar:GeneralRGequations}. Mathematically, these assumptions mean
\begin{subequations}
\begin{eqnarray}
\beta_t &=& t f_t(\vert \gamma_c t \vert) ,\\
\beta_{\gamma_s} &=& -2t^{-1} (1 + \gamma_s)  f_{\gamma_s}(\vert \gamma_c t \vert) , \label{eq:phenomRGgammas}\\
\beta_{\gamma_c} &=& -2 \gamma_c^2 f_{\gamma_c}(\vert \gamma_c t \vert),
\end{eqnarray}
\end{subequations}
with interpolating functions
\begin{subequations}
\begin{eqnarray}
f_t(x) &\simeq &  \begin{cases}  x, & x \ll 1, \\ \frac{\alpha_t}{x^{\eta_t}}, &  x \gg 1 ,\end{cases} \\
f_{\gamma_s}(x) &\simeq & \begin{cases} x^2, &  x \ll 1 ,\\ \frac{\alpha_s}{x^{\eta_s}}&   x \gg 1 ,\end{cases} \\
f_{\gamma_c}(x) &\simeq & \begin{cases} 1, &   x \ll 1, \\ \alpha_c &   x \gg 1. \end{cases} 
\end{eqnarray}
\end{subequations}
In these equations, the positive phenomenological coefficients $\alpha_{t,s,c}$ and $\eta_{t,s}$ can be expected to be of order one. Note, that $\mathcal F$-invariance\cite{PruiskenBaranovSkoric1999} dictates the prefactor $(1+\gamma_s)$ in Eq.~\eqref{eq:phenomRGgammas}. 
				
Under the additional assumption of positivity of $f_{\gamma_c}(x)$ it follows that ($L_{T_X} < L_T < L_{T_c}$)
\begin{equation}
\min(\alpha_c,1)\frac{{T_c}}{T-T_c} < \vert \gamma_c(L_T) \vert < \max(\alpha_c,1)\frac{{T_c}}{T-T_c}.
\end{equation}
Hence, the relative difference between $T_X$ and $T_c$  is small as
\begin{equation}
\frac{T_X-T_c}{{T_c}} \sim Gi_X = \frac{7 \zeta(3) }{\pi^3 g_X} \ll 1. \label{eq:NLSM:TrueGiNumber}
\end{equation}
Here we use the symbol $Gi_X$ for the Ginzburg-Levanyuk number, Eq.~\eqref{eq:intro:BareGiNumber}, evaluated with the normal state conductance $g_X = g(L_{T_X}) = 2/[\pi t(L_{T_X})] \gg 1$. 			
				
We can further use that $f_{t} (x)$ and $f_{\gamma_s}(x)$ have a maximum $f_{t,\gamma_s}(x) \leq f_{t,\gamma_s}^{\rm max} \sim 1$. Therefore we can estimate
\begin{equation}
\frac{d \ln t}{d (\gamma_c^{-1})} \leq \frac{f_t^{\rm max}}{2\min (\alpha_c,1)},
\end{equation}
which yields
\begin{equation}
\frac{t(L_{T_c}) - t(L_{T_X})}{t(L_{T_c})} \lesssim Gi_X.
\end{equation}
In a similar way find
\begin{equation}
\frac{d \ln Z_\omega}{d (\gamma_c^{-1})} \leq \frac{f_{\gamma_s}^{\rm max}}{t(L_{T_X}) \min (\alpha_c,1)},
\end{equation}
and therefore
\begin{equation}
\frac{Z_\omega(L_{T_c}) - Z_\omega(L_{T_X})}{Z_\omega(L_{T_c})} \lesssim Gi_X^2.
\end{equation}
			
All in all, we have determined the critical temperature of the fermionic system $T_c \sim T_X$ with accuracy of $Gi_X$ to be the following function of bare parameters
\begin{equation}
\frac{T_X}{T_{\rm BCS}} =  \exp\left(-2 y_X - \frac{1}{\gamma_c(l)}\right). \label{eq:finalTX}
\end{equation}
Here $y_X$ is the running RG logarithmic scale $y$ at which the system reaches the condition \eqref{eq:LX}.

\subsection{Ginzburg-Landau functional}
\label{sec:GLtheory}

In Sec.~\ref{sec:NLSMRG} we discussed the range of length scales $L \in [l,L_{T_c}]$ and found that the normal-state NL$\sigma$M RG can be applied to determine the critical temperature $T_c$ up to an uncertainty $\sim Gi_X \ll 1$ associated with the critical Ginzburg region. Now we are going to ascend to the next level in our hierarchy of length scales and derive the GL functional.

While Secs.~\ref{sec:GLtheory:saddlepointderiv} and \ref{sec:GL:intrareplica} are devoted to the derivation of the standard GL functional (to all orders in the order parameter field), in Sec.~\ref{sec:GL:interreplica} we go beyond the standard paradigm and derive mesoscopic fluctuations\cite{SkvortsovFeigelman2005} of the parameters in the GL expansion. The perturbative renormalization (Sec.~\ref{sec:Tstar:GL:PertTheory}) of the GL functional incorporates the effect of thermal order-parameter fluctuations in our theory.

\subsubsection{The intrareplica GL functional}
\label{sec:GLtheory:saddlepointderiv}

This derivation of the GL functional for the static order-parameter field $\Delta_\alpha( \v x)$ is based on the normal state NL$\sigma$M at scale $L_{T_X}$ using (i) the local solution of the gap equation, Eq.~\eqref{eq:Tstar:Lambdabar}, and (ii) the implicit definition of the critical temperature $T_c$ via $\gamma_c^{-1}(L_{T_X}) \equiv \ln(T_c/T_X) $.\cite{footnoteTc}
In view of (i), we perform the same rotation, Eq.~\eqref{eq:Tstar:NLSM:BogoliubovTrafo}, as above, but with spatially dependend order parameter field $\Delta_\alpha(\v x)$. After this, we evaluate the resulting NL$\sigma$M action at $q = \Lambda$, which yields the sought GL functional.

The frequency term and the quadratic Hubbard-Stratonovich term generate the exact GL potential
\begin{subequations}
\begin{eqnarray}
S^{\rm GL}_{\rm pot} &=&\int_{\v x} \left [ - 2Z_\omega \tr[q \hat \varepsilon] - \frac{4Z_\omega}{\pi \gamma_c T} \sum_{\alpha} \vert \Delta_\alpha\vert^2 \right ]_{q = \Lambda} \notag \\
&=&- 2Z_\omega  \sum_{\alpha} \int_{\v x} \left \lbrace 8 \sum_{n_1  =0}^\infty \sqrt{\epsilon_{n_1}^2 + \vert \Delta_\alpha \vert^2} + \frac{2}{\pi \gamma_c T} \vert \Delta_\alpha \vert^2\right  \rbrace \notag \\
&\doteq & 16Z_\omega\pi T\sum_\alpha\int_{\v x} \left \lbrace \vert \delta_\alpha \vert ^2 \frac{\ln(T/T_c)}{4} \right.\notag \\
&&\left.+ \sum_{l = 2}^\infty \frac{\left (-1 \right )^l (2l)! (2^{2l - 1}-1)}{(2l-1)(l!)^2 4^l 2^{2l - 1}} \zeta(2l-1)\vert \delta_\alpha \vert ^{2l} \right \rbrace \label{eq:Tstar:GL:GLpotential}
\end{eqnarray}
We introduced the reduced order parameter field $\delta_\alpha = \Delta_\alpha /( \pi T)$ and used ingredient (ii).  At the equality sign with a dot $\doteq$ the divergent Matsubara sum was cut at $\epsilon_n = T_X$.
%
%
In a similar way, we evaluate the kinetic term at the saddle point level:
\begin{eqnarray}
S^{\rm GL}_{\rm kin} &=& \frac{g}{32} \Tr \left [(\nabla [U_\Delta(\v x)^\dagger q U_\Delta(\v x)])^2\right ]_{q = \Lambda} \notag \\
&=&\frac{g}{16}  \sum_{\alpha} \int_{\v x}\left \lbrace \vert \partial_i \Delta_\alpha \vert ^2 \frac{\tanh (\vert \Delta_\alpha \vert/2T)}{\vert \Delta_\alpha \vert T}\right. \notag \\
&-&\left. \frac{ [ \partial_i \vert \Delta_\alpha \vert ^2 ]^2[- \vert \Delta_\alpha \vert + T \sinh( \vert \Delta_\alpha \vert /T)]}{8 \vert \Delta_\alpha \vert^3 T^2 \left [1 + \cosh(\vert \Delta_\alpha \vert/T)\right ]}\right \rbrace . \label{eq:Tstar:GL:kineticTerm}
\end{eqnarray}
\label{eq:Tstar:GL:FullFunctional}
\end{subequations}

In view of the fact, that we are interested in temperatures parametrically close to the mean field temperature $T \sim T_\bkt \sim T_{\rm MF}$, we will keep only leading terms in the GL expansion:
\begin{equation}
S = \frac{\mathcal F}{T} = \frac{1}{T} \sum_\alpha \int_{\v x} A \vert \Delta_\alpha  \vert^2 + \frac{B}{2} \vert \Delta_\alpha  \vert ^4 + C \vert \nabla\Delta_\alpha  \vert ^2, \label{eq:GL:StdFunctional}
\end{equation}
where
\begin{subequations}
\begin{eqnarray}
A &=& \frac{4Z_\omega}{\pi }\ln \frac{T}{T_c} , \\
B &=&Z_\omega\frac{7 \zeta (3)}{2 \pi^3 T^2} ,  \\
C &=& \frac{g}{32T} .
\end{eqnarray}
\label{eq:GL:parametersGL}
\end{subequations}
If renormalizations are discarded, we have $g = g_D$ and $Z_\omega = \pi \nu/4$, and the parameters of the GL functional take the standard form,\cite{AbrikosovGorkovDzyaloshinski1963,LarkinVarlamov2005} $A   = \nu \ln (T/T_c)$, 
$B  = 7 \zeta(3)\nu / (8\pi^2 T^2)$ and $C = \pi \nu D(l)/(8T)$.

\subsubsection{Fluctuation corrections: Intrareplica terms}
\label{sec:GL:intrareplica}

In Sec.~\ref{sec:GLtheory:saddlepointderiv} the GL functional was derived using the local saddle point solution of the NL$\sigma$M. However, more precisely, the GL functional is determined by 
\begin{equation}
S^{\rm GL}[\Delta_{\alpha}] := -\ln \left [\int \mathcal D Q \prod_{n \neq 0, \alpha}\mathcal D [ \Delta_{\alpha,n}, \bar \Delta_{\alpha,n}] e^{-S_{\rm HS}[Q,\check \Delta]}\right ], \label{eq:Tstar:GL:FormalDefSGL}
\end{equation}
where $S_{\rm HS} [Q,\check \Delta]$ was introduced in Eq.~\eqref{eq:NLSM:SSBNLSM:DecoupledNLSM} and $\Delta_{\alpha,n = 0} ( \v x) \equiv \Delta_{\alpha}(\v x)$. The functional integral is dominated by the local saddle point configurations leading to Eq.~\eqref{eq:Tstar:GL:FullFunctional}, while corrections are generated by fluctuations around these mean-field solutions. Generally, such terms lead to a renormalization of GL parameters $A,B,C$ in Eq.~\eqref{eq:GL:parametersGL}. These effects are already taken into account in our approach: the solution of exact RG equations~\eqref{eq:Tstar:GeneralRGequations} includes all corrections stemming from scales $L \in [l, L_{T_c}]$.
As far as corrections from larger scales are concerned, these will be discussed in Sec.~\ref{sec:Tstar:GL:PertTheory} in the framework of the perturbative treatment of the GL functional. We will discard, however, weak (anti-) localization corrections to $C$  (stemming from scales between $L_{T_c} \approx L_T$ and the dephasing length $L_\phi$). Indeed, these corrections are of the order
\begin{equation}
\frac{\delta C^{\rm WL}}{C} \sim Gi_X \ln Gi_X
\end{equation}
and turn out to be subleading with respect to corrections discussed in Sec.~\ref{sec:Tstar:GL:PertTheory}. 

\subsubsection{Fluctuation corrections: Disorder terms}
\label{sec:GL:interreplica}

As shown in Appendix~\ref{App:GLDeriv}, the fluctuation determinant associated to linear deviations from the local saddle point solution further leads to qualitatively new terms in the GL functional: these are interreplica interaction terms. As we explain below, most of them can be interpreted as random fluctuations of GL parameters $A$ and $C$. The leading terms in the expansion $\vert \Delta \vert/T$ and under the assumption that fluctuations of $\Delta$ are smooth on the scale $L_T$ are the following:
\begin{eqnarray}
\delta S^{\rm GL}_{\rm dis} &=& - \frac{1}{2T^2}  \sum_{\alpha \beta} \int_{\v x, \v x'} \Big \lbrace \langle \langle A(\v x) A(\v x') \rangle \rangle  \vert \Delta_\alpha \vert ^2_{\v x} \vert \Delta_\beta \vert ^2_{\v x'} \notag \\
&&  + 2\langle \langle A(\v x) C(\v x') \rangle \rangle  \vert \partial_i \Delta_\alpha \vert ^2_{\v x} \vert \Delta_\beta \vert ^2_{\v x'} \notag \\
&& + \frac{\langle \langle C(\v x) C(\v x') \rangle \rangle}{2} \Big [ 2 \vert \partial_i \Delta_\alpha \vert ^2_{\v x} \vert \partial_i \Delta_\beta \vert ^2_{\v x'} \notag \\
&& +(\partial_i \Delta^*_\alpha [s_x]_{ii'}  \partial_{i'} \Delta_\alpha )_{\v x} (\partial_j \Delta_\beta^* [s_x]_{jj'}   \partial_{j'} \Delta_\beta )_{\v x'} \notag \\
&&+(\partial_i \Delta^*_\alpha [s_z]_{ii'}  \partial_{i'} \Delta_\alpha )_{\v x} (\partial_j \Delta_\beta^* [s_z]_{jj'}   \partial_{j'} \Delta_\beta )_{\v x'}  \Big ]
\Big \rbrace \notag \\
\label{eq:GL:CollectionInterreplica}
\end{eqnarray}
The matrices $s_{x,z}$ are Pauli matrices in the space of spatial coordinates $i,i',j,j' \in \lbrace x,y \rbrace$.
The correlations functions of GL parameters entering Eq.~\eqref{eq:GL:CollectionInterreplica} are given by
\begin{align}
&\left\langle\left \langle \left (\begin{array}{c}
A( \v x) \\ 
\frac{2 \pi T}{D} C( \v x)
\end{array} \right ) ^T \left (\begin{array}{c}
A( \v x') \\ 
\frac{2 \pi T}{D} C( \v x')
\end{array} \right ) 
 \right \rangle \right \rangle \notag \\
 &= \frac{\bm{\beta}}{D (2\pi)^3 T} \left (\begin{array}{cc}
\gamma_{AA} & - \gamma_{AC} \\ 
- \gamma_{AC}&\gamma_{CC}
\end{array} \right ) \delta(\v x - \v x') \label{eq:GL:MatrixCorrelfunction}
\end{align}
 with
\begin{subequations}
\begin{eqnarray}
\gamma_{AA}&=& \frac{7 \zeta(3)}{4\pi }, \label{eq:GLmaintext:FluctuationTC:FT}\\
\gamma_{AC}&=& \frac{\pi^3}{64 },   \label{eq:GLmaintext:MixedFluct:FT} \\
\gamma_{CC}&=&  \frac{7 \pi^2 \zeta(3) - 62 \zeta(5)}{8 \pi }. \label{eq:GLmaintext:ConductivityFluct:FT} 
\end{eqnarray}
\label{eq:GL:CollectionCorrelators}
\end{subequations}
We recall that the diffusion coefficient (at scale $L_{T_X}$) is $D = g/16 Z_\omega$. Further, the parameter $\bm{\beta}$ takes values $\bm{\beta} =  4$ in the presence of spin-rotation invariance and $\bm{\beta} = 1$ in the absence of the latter. 

Here we keep only the Gaussian white noise part of the distribution functions for random quantities $A$ and $C$. The omission of interaction corrections\cite{SkvortsovFeigelman2005} beyond the renormalization of $D$ and of higher moments is justified in the limit $Gi_X \ll 1$. A more detailed study is relegated to a future publication. In the given approximation, our result for random fluctuation of GL parameter $A$ is consistent with the result presented in Ref.~\onlinecite{SkvortsovFeigelman2005}. Note that fluctuations of $A$ can be both interpreted as fluctuations of $T_c$ and of the density of states. 
The spatial delta function is smoothened on the scale $L_T$, and in Appendix \ref{App:GLDeriv} we present the exact correlation function in momentum space. 

The GL functional with disorder-induced corrections also contains random fluctuations of higher angular harmonics, see the last two lines of Eq.~\eqref{eq:GL:CollectionInterreplica}. Since we assumed spin singlet superconductivity from the outset, the leading corrections are of $d$-wave type. Their distribution function has zero mean while the Gaussian correlation happens to be half as large as the one for the GL parameter $C$.

\subsubsection{Perturbative renormalization}
\label{sec:Tstar:GL:PertTheory}

We have now completed the derivation of the disordered GL functional, which is the appropriate theory at length scales larger than $L_{T_c}$. We turn now to the role of fluctuations on scales $L \in [L_{T_c}, \xi]$. To investigate their effect, we will employ a self-consistent perturbative treatment. 

We observe that the mean free path of free Cooper pairs scattering off fluctuations of the transition temperature is 
\begin{equation}
l_{\rm bosons} \sim \xi \vert \epsilon_c \vert g^2 \gg \xi,
\end{equation}
where the reduced temperature is
\begin{equation}
\epsilon_c = \frac{T-T_c}{T_c}.
\end{equation}
Therefore, in the considered regime of length scales, the Cooper-pair kinetics is ballistic.

Before turning to self-consistency, we present the simple perturbative correction to the mass term

\begin{equation}
A \rightarrow A\left [1 + 2\frac{Gi_X}{\epsilon_c} \ln \left (\frac{1}{\vert \epsilon_c \vert}\right ) \right ] =\tilde A + 2A \frac{Gi_X}{\epsilon_c} \ln \left (\frac{Gi_X}{\vert \epsilon_c \vert} \right )  \label{eq:GL:perturb:correctionA}
\end{equation}
with
\begin{equation}
{\tilde A} = {\frac{4Z_\omega}{\pi }\left [\frac{T-T_c}{T_c} + 2{Gi_X} \ln \left (\frac{1}{Gi_X}\right ) \right]}.
\end{equation}
This correction stems from quartic term of the clean GL functional, Eq.~\eqref{eq:GL:StdFunctional}. The analogous diagram from the quartic interreplica term, Eq.~\eqref{eq:GL:CollectionInterreplica}, is smaller by additional factor of $Gi_X$.
We note that ${Gi_X}/{\epsilon_c}$ is the same as the parameter $\gamma_c(L_{T_X}) t(L_{T_X})$ from the analysis of NL$\sigma$M RG, Sec.~\ref{sec:NLSMRG}. In the second line we disentangled effects of the two dimensionless parameters in the theory (${Gi_X}/{\epsilon_c}$ and $Gi_X$). The effect of fluctuations is a reduction of the transition temperature, which turns out to be\cite{LarkinVarlamov2005}
\begin{equation}
 T_{\rm MF} = T_c \left (1   - 2 Gi_X \vert \ln Gi_X \vert \right ). \label{eq:GL:TMF}
 \end{equation} 
Further, the shift of transition temperature is taken into account self-consistently. This results in replacing $\epsilon_c \rightarrow \epsilon$, where 
\begin{equation}
\epsilon = (T-T_{\rm MF})/T_{\rm MF},
\end{equation} 
and $A \rightarrow \tilde A$ in the last term of Eq.~\eqref{eq:GL:perturb:correctionA}. 
Summing up all of these effects leads to a replacement
\begin{equation}
A \rightarrow \tilde A \left [1 + 2\frac{Gi_X}{\epsilon} \ln \left (\frac{Gi_X}{\vert \epsilon \vert}\right ) \right ]
\label{eq:A-renormalized}
\end{equation}
in the GL functional, Eq.~\eqref{eq:GL:StdFunctional}. The factor in square brackets in Eq.~\eqref{eq:A-renormalized} can be absorbed into a redefinition of the order parameter field $\Delta_\alpha$. Then, the self-consistent perturbative treatment yields a GL functional of the form of Eq.~\eqref{eq:GL:StdFunctional} and Eq.~\eqref{eq:GL:CollectionInterreplica} with renormalized coefficients
\begin{subequations}
\begin{eqnarray}
A &\rightarrow & \tilde A = \frac{4Z_\omega}{\pi } \epsilon ,\\
B &\rightarrow & \tilde B = \frac{B}{ \left [1 + 2\frac{Gi_X}{\epsilon} \ln \left (\frac{Gi_X}{\vert \epsilon \vert}\right ) \right ]^2} ,\\
C &\rightarrow & \tilde C = \frac{C}{ \left [1 + 2\frac{Gi_X}{\epsilon} \ln \left (\frac{Gi_X}{\vert \epsilon \vert}\right ) \right ] } .
\end{eqnarray}
\label{eq:renormalizedFluct}
\end{subequations}
The parameteres $\gamma_{AA}, \gamma_{AC}$ and $\gamma_{CC}$ are renormalized analogously to the coefficient $B$.

\subsubsection{Metallic side of the mean field transition}

We note that the procedure employed up to now is equally applicable for $T>T_{\rm MF}$ and $T<T_{\rm MF}$. Let us briefly concentrate on the region of weak fluctuations on the metallic side of the superconducting transition, i.e.  $ \epsilon \gg Gi_X$. As explained in Ref.~\onlinecite{BurmistrovGornyiMirlin2015}, in this case the RG has to be stopped at the scale $L_T$. The corrections to the kinetic GL parameter $C \rightarrow \tilde C$ can be associated to the conductance 
\begin{equation}
g \rightarrow g \left [ 1 -  2\frac{Gi_T}{\epsilon} \ln \left (\frac{Gi_T}{\vert \epsilon \vert}\right )\right ]. \label{eq:GL:physicalconductance}
\end{equation}
Here, $Gi_T$ is the Ginzburg-Levanyuk number with $g(L_T)$ evaluated at scale $L_T$. The expression~\eqref{eq:GL:physicalconductance} can be interpreted as the (physical) conductance above $T_{\rm MF}$ and is of the same form as was presented in Eq.~(45) of Ref.~\onlinecite{BurmistrovGornyiMirlin2015}, provided $\ln (L_\phi/L_T) \sim - \ln  ({Gi_X}/{\vert \epsilon \vert} )$.
 
Furthermore, we conclude that the GL coherence length, Eq.~\eqref{eq:IntroGLCorrellength}, is more accurately given by
\begin{equation}
\xi_{\rm GL} (T) \sim  \frac{\xi(0)}{\sqrt{\epsilon}} \sqrt{\frac{D(L_T)}{D(L_{T_c})}}, \quad \epsilon \gg Gi_T. \label{eq:GLxiwithcorr}
\end{equation}
The zero temperature coherence length is $\xi(0) \sim \sqrt{D(L_{T_c})/T_c}$.

\subsection{Disordered O(2) model, superconducting density and vortex unbinding transition}
\label{sec:O2model}

In the previous section we derived the effective GL functional and incorporated the effect of fluctuations from scales $L \in [L_{T_c}, \xi]$. We now reach the largest scale at which a crossover of theories takes place, namely the GL coherence length $\xi  = \sqrt{\tilde C/(2 \vert \tilde A \vert)}$.  In order to obtain an effective description in terms of the $\mathbf U(1)$ NL$\sigma$M similar to Eq.~\eqref{eq:intro:O2model}, we need to determine the mean-field order parameter.

\subsubsection{Order parameter fluctuations}
\label{sec:O2model:OPfluct}

The mean-field equation for the order parameter field is 
\begin{equation}
- \nabla (\tilde C \nabla \Delta_{\rm MF}(\v x)) + \tilde A \Delta_{\rm MF}(\v x) + \tilde B \vert \Delta_{\rm MF}(\v x) \vert^2 \Delta_{\rm MF}(\v x) = 0.
\end{equation}
Here, the quantities $\tilde C( \v x) = \underline{ \tilde C }+ \delta \tilde C(\v x)$ and  $\tilde A( \v x) = \underline{ \tilde A} + \delta \tilde A(\v x)$  fluctuate randomly according to Eqs.~\eqref{eq:GL:MatrixCorrelfunction} and~\eqref{eq:renormalizedFluct}. Iterative, perturbative solution of this equation leads to $\Delta_{\rm MF} = \bar \Delta_{\rm MF} + \delta \Delta_{\rm MF}(\v x)$
\begin{subequations}
\begin{equation}
\vert \bar \Delta_{\rm MF} \vert= \sqrt{-\frac{\underline{\tilde A}}{\tilde B}} = 2\pi T_{\rm MF} \sqrt{\frac{2 \vert \epsilon \vert}{7 \zeta(3)}} \left [1 + 2\frac{Gi_X}{\epsilon} \ln \frac{Gi_X}{\vert \epsilon \vert}\right ],
\end{equation}
and
\begin{equation}
\delta \Delta_{\rm MF} (\v x) =- \bar \Delta_{\rm MF} \int_{\v x', \v k}\frac{\delta \tilde A(\v x')e^{i \v k (\v x - \v x')}}{\underline{\tilde C}[ \v k^2 + \xi^{-2}]}.
\end{equation}
\end{subequations}
Therefore, the order parameter field fluctuates\cite{LarkinOvchinnikov1971,SkvortsovFeigelman2005} on the scale of the coherence length
\begin{eqnarray}
\frac{\langle \langle \Delta_{\rm MF}(\v x)  \Delta_{\rm MF}(\v x') \rangle \rangle}{\bar \Delta_{\rm MF}^2} &=&\frac{\langle \langle \Delta_{\rm MF}(\v x)  \Delta^*_{\rm MF}(\v x')\rangle \rangle}{\vert \bar \Delta_{\rm MF} \vert^2}  \notag \\
= \frac{\langle \langle \Delta^*_{\rm MF}(\v x)  \Delta^*_{\rm MF}(\v x') \rangle \rangle}{[\bar \Delta_{\rm MF}^*]^2}&=& \mathcal K_\Delta(\v x- \v x')
\end{eqnarray}
with Gaussian correlation function
\begin{eqnarray}
\mathcal K_\Delta(\v x)  &=& \frac{\bm{\beta} \tilde \gamma_{AA}}{\underline{\tilde C}^2D (2\pi)^3 T} \int_{\v k} \frac{e^{i \v k \v x}}{\left [\v k^2 + \xi^{-2}\right ]^2 }\notag \\
 &=& \frac{\bm{\beta} \tilde \gamma_{AA}}{\underline{\tilde C}^2D (2\pi)^3 T} \frac{\xi^2}{4\pi} K_{1}\left  (\frac{\vert \v x \vert}{\xi}\right ) \frac{\vert \v x\vert}{\xi} \notag \\
&{\simeq} &\frac{ \bm{\beta}\left . \tilde \gamma_{AA}\right. \pi^6}{49 \zeta^2(3)} \left [\frac{Gi_X}{\vert \epsilon \vert} \right ]^2 \frac{D \delta( \v x- \v x')}{2\pi T}. 
\label{eq:K-Delta}
\end{eqnarray}
where $K_1$ is the first modified Bessel function. The white-noise approximation in the last line of Eq.~\eqref{eq:K-Delta} is justified when the physics on spatial scales much larger than $\xi$ is considered.

It should be noted, that the inclusion of higher Matsubara harmonics can lead to long range $\ln^2(\v k \xi)$ tails in the Fourier transform of $\mathcal K_\Delta(\v x)$.\cite{LarkinOvchinnikov1971} Such effects are beyond our accuracy in the determination of $T_\bkt$, see the discussion in Sec.~\ref{sec:O2model:discussion}.  Further, we explicitly checked that fluctuations of $\tilde C(\v x)$ do not contribute to fluctuations of $\Delta_{\rm MF}(\v x)$.

\subsubsection{Mean field stiffness}

Finally, we find the mean field stiffness, which fluctuates on the scale of $\xi$, to be
\begin{equation}
\frac{K(\v x)}{\pi} = \frac{{\tilde C(\v x)}}{T} \vert \Delta_{\rm MF}(\v x)\vert^2 \approx \frac{\bar K + \delta K(\v x)}{\pi} .
\end{equation}
Following the notation from above, $\bar K$ denotes the average stiffness and $\delta K(\v x)$ the small fluctuations. Replacing $\tilde C$ and $\Delta_{\rm MF}$ by their respective expectation values leads to 
\begin{eqnarray}
\frac{\bar K}{\pi} &=& \frac{g}{32} \frac{\vert \bar \Delta_{\rm MF}^{\rm (0)} \vert^2}{T^2}  \left [ 1 + 2\frac{Gi_X}{\epsilon} \ln \left (\frac{Gi_X}{\vert \epsilon \vert}\right )\right ] \notag \\
&=& \frac{1}{4\pi} \frac{\vert \epsilon \vert}{Gi_X}\left [ 1 + 2\frac{Gi_X}{\epsilon} \ln \left (\frac{Gi_X}{\vert \epsilon \vert}\right )\right ]. \label{eq:GLmeanstiffness}
\end{eqnarray}
Here we have introduced $\vert \bar \Delta_{\rm MF}^{\rm (0)} \vert^2 = - A/B =  - \epsilon 8 \pi^2 T^2 / 7 \zeta(3)$, the mean field gap without corrections stemming from length scales $L \in [L_{T_c}, \xi]$. 

Thus, the mean superconducting density $\bar K(\epsilon)$ ($\epsilon <0$) is determined by the normal state conductivity at the same distance from the transition (i.e. $g( - \epsilon)$ obtained by reflecting $\epsilon = (T - T_{\rm MF})/T_{\rm MF}$ about the origin), see Eq.~\eqref{eq:GL:physicalconductance}.

The fluctuations of $K$ are given by 
\begin{equation}
\frac{\delta K(\v x)}{\bar{K}} = 
\frac{\delta \tilde{C}(\v x)}{\underline{\tilde{C}}}+
\frac{\delta \Delta_{\rm MF}(\v x)}{{\bar{\Delta}}_{\rm MF}} +\frac{\delta \Delta^\star_{\rm MF}(\v x)}{{\bar{\Delta}}^\star_{\rm MF}}   
\end{equation}

The fluctuations of $K$ are goverened by order parameter fluctuations:
\begin{equation}
\langle \delta K(\v x) \delta K(\v x') \rangle = \frac{\bm{\beta} \gamma_{ AA}  }{4} \left [\frac{\pi^3}{7 \zeta(3)}\right ]^2 \frac{D \delta( \v x- \v x')}{2\pi T} . \label{eq:fluctuatingStiffness1}
\end{equation}

Note that the delta function is smoothened on the scale $\xi$. Mesoscopic fluctuations of the GL parameter C lead to subdominant contributions to  $\langle \delta K(\v x) \delta K(\v x') \rangle$. We explicitly checked, that terms proportional to $\gamma_{AC}$ or $\gamma_{CC}$ are negligble as compared to those proportional to $\gamma_{AA}$ in view of additional small factors of $\vert \epsilon\vert$ and/or $Gi_X$.

\subsubsection{BKT transition temperature}
\label{sec:BKTRG}

We remind the reader that the clean $\mathbf U(1)$ NL$\sigma$M supplemented with topological excitations (vortices) exhibits a phase transition which is driven by the logarithmic intervortex interaction. The latter is encoded in the following RG equations for the mean stiffness $\bar K$ and fugacity $\bar z_v$,
\begin{subequations}
\begin{eqnarray}
&&\frac{d\bar K^{-1}}{d y} =  {\bar z_v^2}, \\
&&\frac{d\bar z_v}{d y} = 2\left (1 - \bar K \right ) \bar z_v,
\end{eqnarray}
\label{eq:standardBKTRG}
\end{subequations}
where $y = \ln(L/\xi)$. The fixed point defining the vortex unbinding transition is given by $(\bar K^*,  \bar z_v^*) = (1,0)$.					

Generally, the effect of disorder on the vortex physics in the $\mathbf U(1)$ NL$\sigma$M is twofold: first, the random stiffness modifies the interaction between vortices and rotationless supercurrents as well as between vortices themselves. Second, since the stiffness determines the core energy $S_{\rm core} =- \ln z_v = \alpha_z K$, fluctuations of the stiffness can lead to pinning of vortices.  The coefficient $\alpha_z$ is a number of order unity in realistic superconductors,\cite{BenfattoCastellaniGiamarchi2009} and the fugacity $z_v$ acquires a log-normal distribution.

In the present case, the effect of disorder on the BKT transition temperature is relatively weak: the quartic interreplica terms are RG irrelevant. In addition, since $K(\v x)$ fluctuates on the scale $\xi$, vortex pinning is negligible for the following reason: The size of a vortex core is of the same scale $\xi$ and by consequence fluctuations largely cancel out. 

Technically, the effect of disorder can be taken into account by partial contraction of the quartic interreplica terms.\cite{Jose1981} It turns out that intrareplica quartic terms (thermal fluctuations from length scales $L > \xi$) dominate upon disorder effects by a factor of $Gi_X^{-1}$. 

Using this observation and simple rescaling arguments\cite{FootnoteKbare} in the formal definition of the bare stiffness we find that $K_{\rm bare}$ depends only on the single parameter $Gi_X/\vert \epsilon \vert$
\begin{subequations}
\begin{equation}
\frac{K_{\rm bare}}{\pi} = \frac{1}{4\pi}\frac{\vert \epsilon \vert}{Gi_X} f_K \left (\frac{Gi_X}{\vert \epsilon \vert}\right ) ,
\end{equation}
with $f_K(x)/x$ being a monotonically decreasing function and
\begin{equation}
f_K\left (x\right ) \simeq  \begin{cases} 1 - 2x \ln x, & \text{for } x \ll 1, \\ \leq x^{\eta_K} \; (\eta_K < 1), & \text{for } x \gg 1. \end{cases}
\end{equation}
\label{eq:fKstiffness}
\end{subequations}
In the limit $z_v \rightarrow 0$, the critical stiffness is $K_* \rightarrow 1$. The transition temperature $T_{\bkt}$ is governed by the solution $\alpha_{\bkt}$ of $f_K  (1/\alpha_\bkt) = 4/\alpha_\bkt$, i.e.
\begin{subequations}
\begin{equation}
\vert \epsilon_\bkt \vert \equiv \frac{T_{\rm MF}- T_\bkt}{T_{\rm MF} }=  \alpha_\bkt Gi_X. \label{eq:Tstarfinal}
\end{equation}
Even though the precise value of $\alpha_\bkt$ remains unknown, the asymptotic terms in the limit $\vert \epsilon \vert \gg Gi_X$ are sufficient to identify $\alpha_\bkt$ as a number of order unity.

In a realistic superconductor close to the transition, the fugacity $\bar z_v \sim \mathcal O(1)$ is finite and thus the system undergoes the phase transition far from the critical end point $(\bar K_*,\bar z_v^*) = (1,0)$. This does not alter, however, the result \eqref{eq:Tstarfinal} with 
$\alpha_\bkt$ understood as an unknown coefficient of order unity.  For completeness we repeat that
\begin{eqnarray}
Gi_X &=&  \frac{7 \zeta(3) }{\pi^3 g_X} \ll 1,\\
T_{\rm MF} &=&  T_c \left (1   - 2 Gi_X \vert \ln Gi_X \vert \right ),
\end{eqnarray}
\label{eq:Tstarfinalalltogether}
\end{subequations}
with $g_X$ being the NL$\sigma$M coupling constant evaluated close to the superconducting instability of the diffusive Fermi liquid occurring at the energy scale $T_c$. The expression for the vortex unbinding temperature, defined via Eq.~\eqref{eq:Tstarfinal}, is one of the major results of this paper.

We speculate now on the behavior of the vortex unbinding temperature, when the system is tuned close to the quantum critical point [characterized by $(\gamma_c (L_{T_c} = \infty), t(L_{T_c} = \infty)) = (-\infty, t_{\sit} \sim 1)$] of the NL$\sigma$M RG from Sec.~\ref{sec:NLSM:RGdiscussion}. 
It appears very plausible that Eq.~\eqref{eq:Tstarfinal} describes the first term of a geometric series, such that
\begin{equation}
T_\bkt =  \frac{T_{\rm MF}}{1+ \alpha_\bkt Gi_X }.\label{eq:TstarfinalSpecul}
\end{equation}
A result of this form was first obtained by Beasley, Mooij and Orlando,\cite{BeasleyMooijOrlando1979} who however omitted all fluctuation and disorder renormalizations. (In their approximation $T_{\rm MF} = T_{\rm BCS}$, $ \alpha_\bkt  = 4$, and $Gi_X  = Gi$.) 

\subsubsection{Correlation length}
\label{sec:Correllength}

It is easy to check that the quantity
\begin{equation}
c(K) = 4[K^{-1} - 1 - \ln K^{-1}]- z_v^2
\end{equation}
is conserved under the BKT RG, Eqs.~\eqref{eq:standardBKTRG}. On the normal conducting side, the limit $\tau_\bkt \rightarrow 0$ implies 
\begin{equation}
 c \simeq - \left (\frac{\pi}{2b}\right )^2 \frac{ \tau_\bkt}{2 Gi_X}.
 \label{eq:Assumptionc}
 \end{equation} 
We remind the reader of the definition $\tau_\bkt = (T-T_\bkt)/T_\bkt$. The fitting parameter $b$ is chosen such that the solution of RG Eqs.~\eqref{eq:standardBKTRG} with $c(K)$ given by Eq.~\eqref{eq:Assumptionc} leads to the same form as Eq.~\eqref{eq:IntroBKTCorrellength}
\begin{equation}
 \xi_\bkt(T)\sim \xi(T_\bkt) e^{b\sqrt{\frac{Gi_X}{\tau_\bkt}}},  \quad \tau_\bkt \ll Gi_X
\end{equation}
 for the correlation length.~\cite{Footnote-LBKT} 
  As was discussed in detail in Ref.~\onlinecite{BenfattoCastellaniGiamarchi2009}, the parameter $Gi$ (or better $Gi_X$) naturally appears in the standard definition of both BKT and GL coherence lengths.


\subsubsection{Temperature-dependent stiffness}
\label{sec:stiffnessLowT}

We now focus on the superconducting side of the transition and discuss the temperature dependent mean stiffness. We consider three different temperature regimes: (i) $T \ll T_\bkt$, (ii) $Gi_X \ll |\tau_\bkt| \ll 1 $ and (iii) $|\tau_\bkt|  \ll Gi_X$.

In the immediate vicinity of the transition [regime (iii)], Eq.~\eqref{eq:Assumptionc} implies
\begin{equation}
K(\tau_\bkt) = \frac{1}{1- \frac{\pi}{4b} \sqrt{|\tau_\bkt|/ Gi_X}}. \label{eq:stiffnessnearTBKT}
\end{equation}
In contrast, in regime (ii), renormalization effects are negligible and Eq.~\eqref{eq:fKstiffness} describes the temperature dependent stiffness.

Finally, we consider the regime (i) of lowest temperatures, $T \ll T_\bkt$. It is useful to first clarify the hierarchy of length scales occurring in this regime. As compared to the case $-\tau_\bkt \ll 1$ depicted in Fig.~\ref{fig:lengthscales}, now the following scales are almost identical: $L_{T_c} \sim L_{\Delta} \sim \xi$. Therefore, the large window $L_{T_c} \ll L \ll \xi$ of dominant thermal fluctuations disappears and the stiffness is determined by the outcome of the NL$\sigma$M RG,
\begin{equation}
\frac{K}{\pi} =  \frac{g(L_{T_c})}{8} \frac{\Delta}{2T} \tanh \frac{\Delta}{2T} . \label{eq:stiffnesslowT}
\end{equation}
Within our theory, $\Delta = \Delta(T) \gg T$ is the temperature-dependent spectral gap, which has standard BCS form, but with renormalized $T_c$, as dictated by the NL$\sigma$M RG.
This expression incorporates the quantum corrections stemming from scales $L \lesssim L_{T_c}$.
The low-temperature stiffness (superconducting density) including dominant quantum corrections to $T_c$ was recently determined in Ref.~\onlinecite{Skvortsov2015} using an alternative approach (self-consistent diagrammatic technique).

\subsection{Summary and discussion}
\label{sec:O2model:discussion}

\begin{figure*}
\begin{minipage}{.45\textwidth}
\includegraphics[scale=.6]{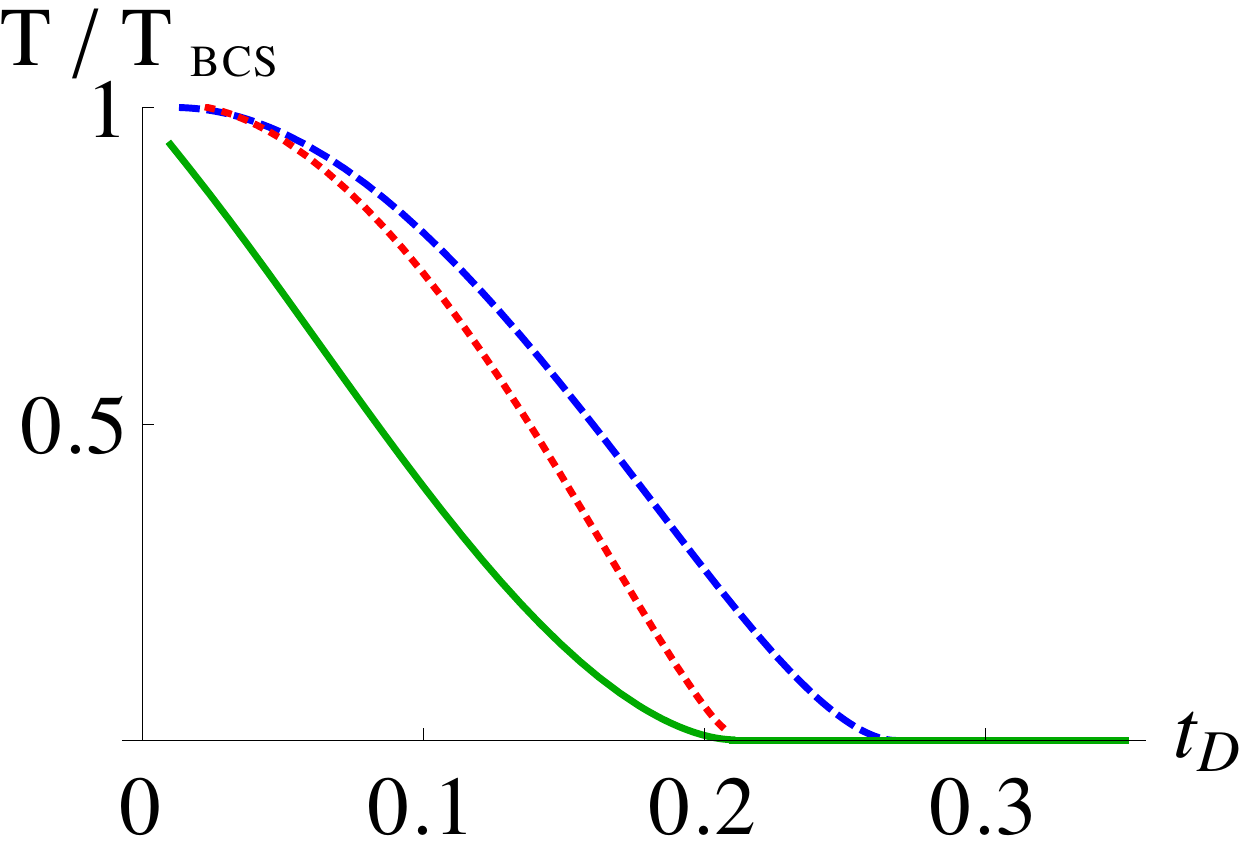}
\end{minipage}
\begin{minipage}{.45\textwidth}
\includegraphics[scale=.6]{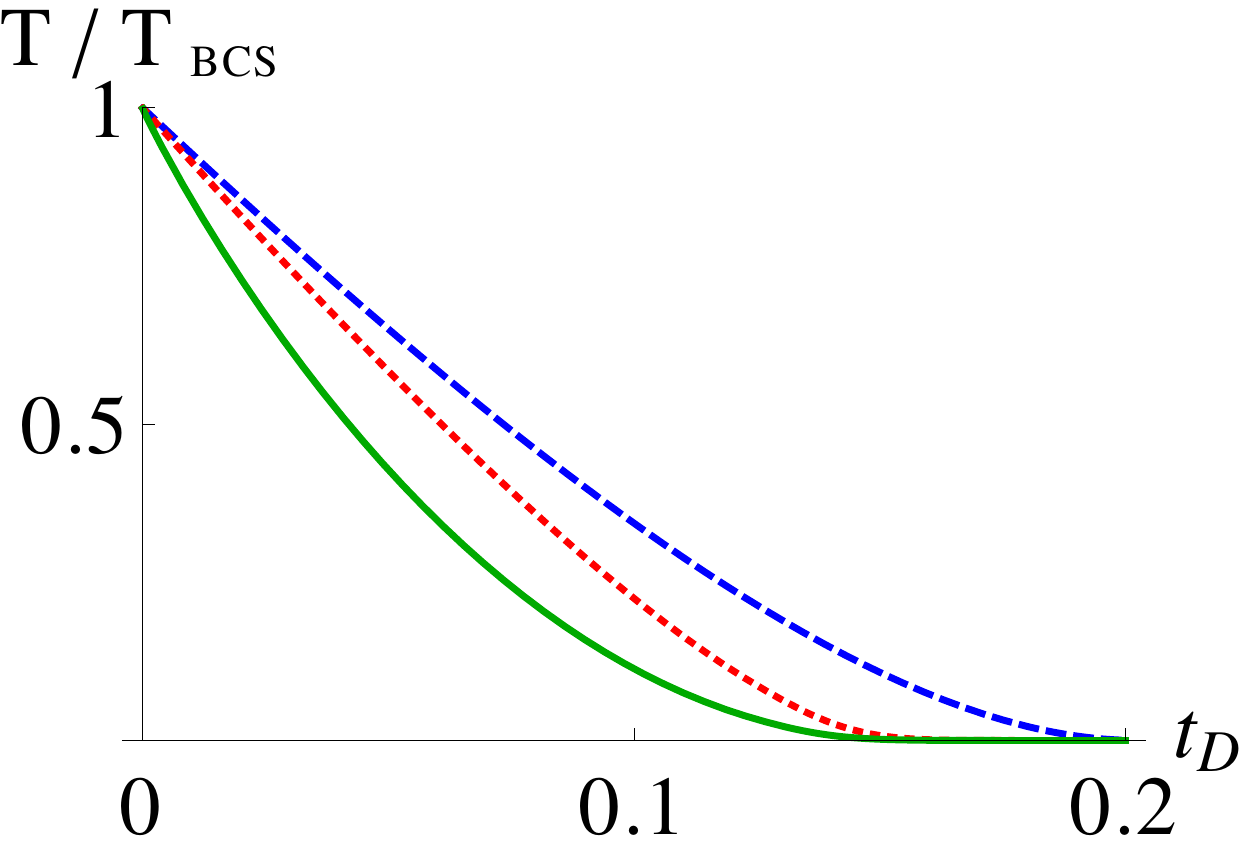}
\end{minipage}
\caption{Transition temperature for a homogeneously disordered superconducting film with Coulomb interaction and strong spin-orbit scattering as a function of the Drude resistance $t_D = 1/(\pi \mu \tau)$. On the left (right), the mean free time $\tau$ is varied (fixed) while the chemical potential is kept fixed (is varied). Blue dashed: Finkelstein's approximate solution for $T_c$, Eq.~\eqref{eq:intro:FinkelsteinSuppresion}. Red dotted: Temperature $T_X$, Eq.~\eqref{eq:finalTX}, of the instability of the fermionic system. Green solid: Vortex unbinding temperature, Eq.~\eqref{eq:TstarfinalSpecul}. Green and red curves are obtained by numerical solution of RG equations. The only free parameter in the left (right) plot is $T_{\rm BCS}/\mu = 0.01$ [$\ln (T_{\rm BCS} \tau )= -0.2$]. The origin of the different behavior in left and right plots is due to the strong dependence of $\gamma_c(l)$ on $t_D$ in the left graph.}
\label{fig:transitiontemperatures}
\end{figure*}

The final result for the vortex unbinding temperature is visualized in Fig.~\ref{fig:transitiontemperatures}. For concreteness, we consider a system with long-range Coulomb interaction and strong spin-orbit coupling. The transition temperature is displayed as a function of the Drude resistance $t_D$. Deviation from the horizontal line, $T/T_{\rm BCS}=1$, implies a violation of Anderson's theorem.  
The three curves correspond to Finkelstein's solution, Eq.~\eqref{eq:intro:FinkelsteinSuppresion} (blue dashed line), to the temperature $T_X$ of the superconducting instability, Eq.~\eqref{eq:finalTX}, as resulting from the full  one-loop fermionic RG  (red dot-dashed line), and to the vortex unbinding temperature $T_{\rm BKT}$, Eq.~\eqref{eq:TstarfinalSpecul} (green full line). We see that there is a substantial difference between the three curves.
The relative difference between $T_X$ and $T_{\rm BKT}$ is of the order of $Gi_X |\ln Gi_X|$, see Eqs. \eqref{eq:NLSM:TrueGiNumber}, \eqref{eq:GL:TMF}, and \eqref{eq:Tstarfinal}. This difference can serve as a rough estimate for the width of the BKT fluctuation regime in the temperature dependence of resistivity, which is of order $Gi_X$, as will be discussed in Sec.~\ref{sec:resistivity}.

Let us emphasise once more that, as seen in Fig.~\ref{fig:transitiontemperatures}, the found mean-field transition temperature $T_{\rm MF}$, and consequently the true transition temperature $T_\bkt$, can differ strongly from the BCS temperature $T_{\rm BCS}$.  The dominant effects behind this difference are those  incorporated in the NL$\sigma$M RG. In addition to them, there is a contribution of GL fluctuation corrections. 
The randomness of the stiffness inherited from ``mesoscopic'' fluctuations of GL parameters, Eq.~\eqref{eq:GL:CollectionCorrelators}, is parametrically smaller and thus of minor importance for the finite-temperature superconducting transition that we are exploring. This applies also to $\ln^2(\v k \xi)$ tails\cite{LarkinOvchinnikov1971} in the correlation function of $K(\v x)$, Sec.~\ref{sec:O2model:OPfluct}, which are subleading and do not affect Eq.~\eqref{eq:Tstarfinal}.

Therefore, our theory implies that in homogeneously disordered films the long-wavelength theory describing the finite temperature vortex unbinding transition is a theory of clean bosons. In this paper we showed how this theory emerges from the underlying theory of disordered fermions. Our derivation of parameters of the effective $\mathbf{U}(1)$ theory takes into account all essential disorder, quantum-interference, and interaction effects. 

It is worth pointing out that we have assumed a short-range disorder. Spatial variations of the chemical potential and/or the BCS-coupling constant with a characteristic length large compared to $\xi(T_\bkt)$ can be incorporated in the resulting bosonic theory, which will yield
a disordered $\mathbf U(1)$ NL$\sigma$M. This will lead to an additional suppression of $T_\bkt$.\cite{BenfattoCastellaniGiamarchi2009} 

\section{Temperature dependence of resistance}
\label{sec:resistivity}

This section is devoted to the temperature dependence of the resistance in a 2D metallic film at the verge of superconductivity.

\subsection{Interpolating function for the resistance}

The asymptotic behavior of the resistance in the vicinity of the BKT transition is given by generalization of the result obtained in Ref.~\onlinecite{HalperinNelson1979}. 
\begin{equation}
\rho(T) \simeq A_v^{-1} {t_X} \left (\frac{\xi(T_\bkt)}{\xi_{\hn}(T)}\right )^2,   \qquad  \tau_\bkt\ll Gi_X
\label{eq:rho1}
\end{equation}
and by Eq.~\eqref{eq:GL:physicalconductance}
\begin{eqnarray}
\rho(T) &\simeq& t(L_T) \left [1 +4\left (\frac{\xi_{\hn}(T)}{\xi(T_\bkt)}\right )^2 \ln \left (\frac{\xi_{\hn}(T)}{\xi(T_\bkt)}\right )\right ], 
\nonumber \\
&& \hspace*{3cm}
\tau_\bkt\gg Gi_T.
\label{eq:rho2}
\end{eqnarray}
Here $\xi_{\hn}$ is a generalization of the Halperin-Nelson length,\cite{HalperinNelson1979} cf. Eq.~\eqref{eq:introHNCorrellength},
\begin{equation}
\xi_\hn(T) = \frac{\xi(T_\bkt)}{b} \sinh \left (b \frac{\tilde {Gi}_T}{\tau_\bkt}\right ),
\end{equation}
inasmuch it contains all renormlization effects discussed in this work. We introduced an interpolating function for the Ginzburg number $\tilde {Gi}_T = 7 \zeta(3) \tilde t(T)/(2\pi^2)$ by means of $\tilde t(T)$ given by  $\tilde t(T)= t(L_T)$ for $T\geq T_X$ and by the extrapolation 
\begin{equation}
\tilde t(T) = T_X Gi_X \partial_T t(L_T)\vert_{T_X} \tanh\left [\frac{T-T_X}{T_X Gi_X}\right ] + t(L_{T_X})
\end{equation}
 for temperatures below $T_X$. 

The vortex-generated resistance in the immediate vicinity of $T_\bkt$, Eq.~\eqref{eq:rho1}, contains an unknown prefactor  $A_v^{-1}$ of order unity. The resistance far from the superconducting transition includes the fluctuation corrections, Eq.~\eqref{eq:GL:physicalconductance}, which are essentially the anomalous Maki-Thompson correction. In 2D, they dominate upon Aslamazov-Larkin corrections. We propose the following interpolating function for the resistance which reproduces correctly both limits \eqref{eq:rho1} and \eqref{eq:rho2}:
 \begin{equation}
{\rho(T)} = \frac{\tilde t(T)}{1 + 2 \left (\frac{\xi_{\hn}(T)}{\xi(T_\bkt)}\right )^2 \ln \left [e^{A_v/2} + \left (\frac{\xi(T_\bkt)}{\xi_{\hn}(T)}\right )^2\right ]}. \label{eq:resistanceIpolfunction}
\end{equation}

Let us comment on the width of the crossover region from metal to superconductor in the temperature dependence of resistance. As is indicated by Eqs.~\eqref{eq:rho1} and \eqref{eq:rho2}, our formalism predicts that this crossover happens within a window of relative size $Gi_X$ around the mean field temperature $T_{\rm MF}$. The only temperature scale arising in our analysis of the crossover is  $\vert \epsilon\vert \sim {Gi_X}$.

\subsection{Electrical vs. heat transport}

\begin{figure}
\includegraphics[scale=.4]{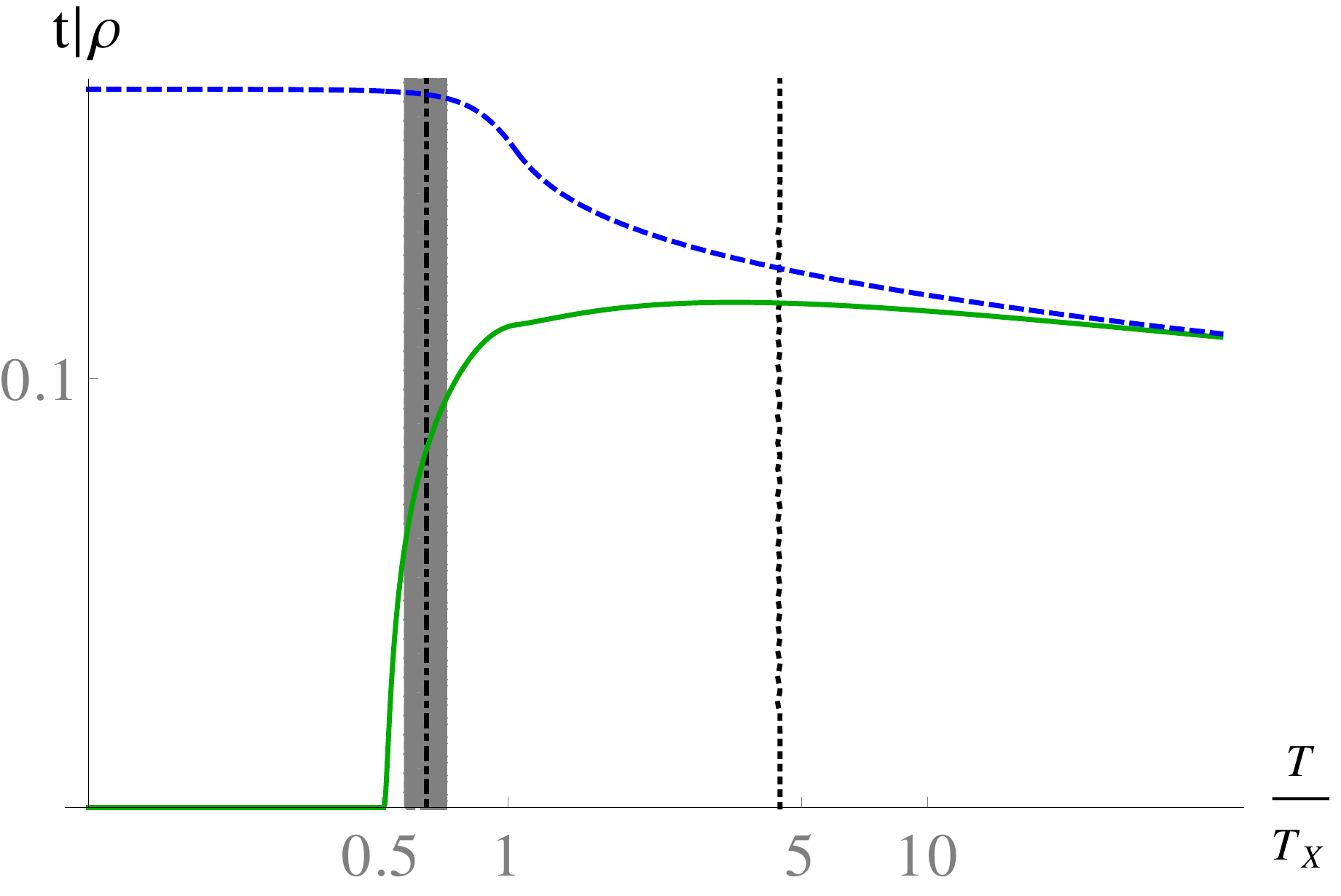} 
\caption{Temperature dependent resistance of a homogeneously disordered metallic film close to the superconducting transition. Here, the case of strong spin-orbit coupling and Coulomb interaction is considered. Blue dashed curve: temperature dependence of the NL$\sigma$M coupling constant $t$, which determines the prefactor of Eq.~\eqref{eq:ThermalConductivity} for the thermal conductivity. Green solid curve: temperature dependence of the physical resistance. The BCS transition temperature $T_{\rm BCS}$ is marked by the black dashed vertical line, while the mean field transition temperature $T_{\rm MF}$ is marked by a dot-dashed line surrounded by a gray Ginzburg-Levanyuk window. Note the logarithmic scale of the temperature axis. This plot was generated using Eq.~\eqref{eq:resistanceIpolfunction} with the bare values $\ln (T_{\rm BCS} \tau) = -0.2$ and $t_D = 0.1$, as well as phenomenological constants $\alpha_\bkt = 4$, $A_v = \pi^4/56 \zeta(3)$ and $b = 1$.}
\label{fig:resistance}
\end{figure}

The final result for the resistivity, Eq.~\eqref{eq:resistanceIpolfunction}, is displayed in Fig.~\ref{fig:resistance}. Again we focus on the case of a system with strong spin-orbit coupling and Coulomb interaction.

When the transition is approached, the physical resistance $\rho$ starts to deviate from the NL$\sigma$M running coupling constant $t$ and eventually rapidly decreases to zero. The difference between the two quantities is due to inelastic processes, in particular fluctuation corrections, which are beyond the RG scheme of the NL$\sigma$M. The quantity $t(L)$ serves as a starting point for the calculation of various physical observables subjected to these corrections, including the electrical resistance, Eq.~\eqref{eq:resistanceIpolfunction}, and the superconducting density, Eq.~\eqref{eq:fKstiffness}. We interpret the NL$\sigma$M coupling constant $t(L)$ [or $\tilde t(T)$] as the electronic contribution to the thermal resistance of the system. Indeed, it is known that the quasiparticle contribution to the thermal transport coefficient $\kappa$ has no singular corrections at the onset of the transition.\cite{NivenSmith} Technically, this is related to the extra frequencies at heat-current vertices (see Appendix \ref{app:Kappa}), each producing an extra factor $\gamma_c^{-1}$ in the inelastic contribution to the thermal conductivity as compared to the electrical conductivity (see Ref. \onlinecite{BurmistrovGornyiMirlin2015} for the NL$\sigma$M calculation of the latter). In particular, the thermal conductivity $\kappa$ can be readily derived within our NL$\sigma$M formalism at the saddle-point level (see Appendix \ref{app:Kappa} for the technical details), with the result 
\begin{equation}\label{eq:ThermalConductivity}
\kappa = \frac{gT}{4\pi}\int_{\Delta/T}^\infty \frac{\zeta^2d\zeta}{\cosh^2(\zeta/2)},
\end{equation}
and is thus related to renormalized $g$, and consequently, to renormalized $t$. 

To summarize, well above the transition both electric resistivity  and thermal resistivity $t$ follow the NL$\sigma$M RG. When the transition is approached, the electric resistivity $\rho$ gets strongly suppressed due to fluctuation effects and eventually becomes zero at $T_{\rm BKT}$. On the other hand, the NL$\sigma$M coupling $t$, and thus the thermal resistivity, remains finite in the vicinity of the transition temperature $T_{\rm BKT}$.

It should be mentioned that in the model of diffusive electronic system with long-range Coulomb interactions there are additional logarithmic corrections to the thermal conductance $\kappa$.\cite{WiedemannFranzViolation}
These corrections, however, are nonuniversal and beyond the RG scheme,\cite{SchwieteFinkelstein2014}  so that they are not accounted for explicitly in our treatment.

%

\subsection{Finite size effects and the role of electromagnetic fields}
\label{sec:finitesize}

Finite size effects may lead to a smearing of the BKT transition. First, as for any phase transition, the linear dimension $L_\square$ of the film leads to a natural infrared cut-off. Second, a finite size effect which is specific to superconducting films is due to the finite thickness $d \neq 0$, as has been already mentioned in Sec.~\eqref{sec:normal-nlsm}.
Due to the interplay with electromagnetic fields, a finite thickness leads to an infrared cut-off for the logarithmic vortex-antivortex interaction\cite{Pearl1964} at the length scale of $\lambda^2/d$, where $\lambda$ is the London penetration depth. 
In the context of the present treatment, this means that RG has to be stopped at the smaller of the two scales $\lbrace \lambda^2/d, L_\square\rbrace$, which at the same time is the maximal value for $\xi_{\rm BKT}$. The consequence is a rounding\cite{BenfattoCastellaniGiamarchi2009} of the sharp resistance drop near $T_\bkt$ in Fig.~\ref{fig:resistance}.

\section{Summary and Outlook}
\label{sec:summary}

To summarize, we have developed a theory for the vortex unbinding transition in homogeneously disordered superconducting films. This theory incorporates the effects of quantum, mesoscopic and thermal fluctuations stemming from length scales ranging from the superconducting coherence length down to the Fermi wavelength. Using the developed theory, we determine the dependence of essential observables (including the vortex unbinding temperature, the superconducting density, as well as the temperature-dependent resistivity and thermal conductivity) on microscopic characteristics such as the disorder-induced scattering rate and bare interaction couplings. More specifically, our key results are as follows: 

\begin{enumerate}

\item
We have performed a consecutive mapping of emerging effective field theories, starting from the Fermi-liquid theory at a short scale through the  normal-state interacting sigma model, superconducting-state sigma model, and Ginzburg-Landau theory, to the $\mathbf{ U}(1)$ theory of superconducting phase fluctuations.

\item
As a result of this procedure, we obtained the superconducting density in the temperature regimes (i) $T \ll T_\bkt$, Eq.~\eqref{eq:stiffnesslowT}, and (ii) $Gi_X \ll \vert \tau_\bkt \vert = \vert T-T_\bkt \vert/T_\bkt \ll 1$, Eqs.~\eqref{eq:GLmeanstiffness}, \eqref{eq:fluctuatingStiffness1}, \eqref{eq:fKstiffness}, which include effects from quantum, mesoscopic and thermal fluctuations. We find that in regime (ii) the superconducting density at temperature $T$ is determined by the the normal state conductance at the temperature $2T_{\rm MF} - T$ obtained by reflecting about the mean field transition temperature $T_{\rm MF}$. The scale $L_{T_X}$ and the associated Ginzburg number $Gi_X$ are provided by the fermionic nonlinear sigma model RG.

\item
Combining our result for the superconducting density in regime (ii) with the RG treatment of the BKT transition, we have found the vortex unbinding temperature $T_\bkt$, Eq.~\eqref{eq:TstarfinalSpecul}. Figure~\ref{fig:transitiontemperatures} depicts the transition temperature as a function of experimentally controllable parameters in the case of a system with strong spin-orbit coupling.
Further, we extracted the behavior of the stiffness in the immediate vicinity of the transition $\vert \tau_\bkt \vert \ll Gi_X$, Eq.~\eqref{eq:stiffnessnearTBKT}. We thus provided results for the superconducting density in all regimes below $T_\bkt$.

\item 
We have proposed a function, Eq.~\eqref{eq:resistanceIpolfunction},  which interpolates between vortex-dominated (exponentially small) resistance close to the vortex unbinding temperature and the Maki-Thompson-like fluctuation resistivity representing the dominant fluctuation correction further from the transition. We find, that the resistance drop occurs in a small temperature window of strong fluctuations of relative size $Gi_X$. Our result for the resistance is plotted in Fig.~\ref{fig:resistance} for the case of a system with strong spin-orbit scattering.

\item
We have identified the running charge $t(L)$ of the NL$\sigma$M RG scheme of our theory with the electronic contribution to the thermal resistance of the system, see Eq.~(\ref{eq:ThermalConductivity}).  

\item
We have performed a derivation of mesoscopic fluctuations of Ginzburg-Landau coefficients, including both the mass term and the prefactor of the kinetic term,  for both cases of the preserved and broken spin-rotation invariance.

\end{enumerate}

Before closing the paper, we briefly discuss some of perspective directions for future work.

\begin{enumerate}

\item
Our results, and in particular Eqs.~(\ref{eq:TstarfinalSpecul}) and (\ref{eq:resistanceIpolfunction}), should be useful for the analysis of experimental data on $T_{\mathrm{BKT}}$ and on resisitivity near the transition,
including recent measurements\cite{El-BanaBending2013,ZhaoXue2013,YongSiegel2013,KoushikGhosh2013,GangulyBenfatto2015,Schneider2014,BaturinaVinokur2012} and expected future experiments.
It would be also very interesting to prove experimentally our observation that the distance $(T_\bkt-T_{\rm MF})/T_{\rm MF}$ should be of the order of the dimensionless thermal resistivity.

\item
This paper was devoted to the thermodynamic phase transition into the superconducting state. While renormalizations on intermediate scales were of quantum origin, the physics in the close vicinity of the transition was governed by thermal fluctuations. The corresponding transition line ends at $T=0$ at the point of superconductor-insulator transition,\cite{GantmakherDolgopolov2010} which represents a prominent example of a quantum phase transition and is driven by quantum fluctuations. Contrary to the finite-temperature transition controlled by a Gaussian fixed point, the fixed point governing the quantum superconductor-insulator transition is at strong coupling, which makes its controllable analytical exploration an extremely difficult task. While considerable progress has been achieved within theories of disordered bosons \cite{Efetov1980,FisherFisher1989,ChamonNayak2002} and within the fermionic NL$\sigma$M RG, see Ref.~\onlinecite{BurmistrovGornyiMirlin2015}, development of a controllable theory of the critical behaviour at this quantum phase transition remains an outstanding challenge for future research. 

\end{enumerate}

\section*{Acknowledgements}

We would like to thank A. Finkelstein, N. Kainaris, D. Khveshchenko, M. Scheuer, J. Schmalian, M. Skvortsov, K. Tikhonov, and A. Tsvelik for useful discussions. This work was supported by Deutsche Forschungsgemeinschaft (DFG),
by NSF Grants No. DMR-1401908 and ECCS-1407875, and in part by
DAAD (German Academic Exchange Service) grant (E.J.K. and A.L.),
and by Russian Science Foundation under the grant
No. 14-42-00044 (I.V.P., I.V.G., I.S.B., and A.D.M.).

\appendix

\section{Derivation of disordered GL theory from the NL$\sigma$M}
\label{App:GLDeriv}

In this appendix, the interreplica corrections to the GL functional are derived. To this end, the functional integral Eq.~\eqref{eq:Tstar:GL:FormalDefSGL} is evaluated beyond the saddle point approximation.

Starting point for the derivation of the interreplica terms of the GL functinal is the action~\eqref{eq:NLSM:SSBNLSM:DecoupledNLSM}, which yields, after integration of Hubbard-Stratonovich fields $\Delta_{\alpha, n}$ with $n \neq 0$,
\begin{eqnarray}
S &=& \int_{\v x} \frac{g}{32} \tr (\partial_i Q)^2 - 2Z_\omega\tr \left [Q \left [\hat \epsilon + \left (\begin{array}{cc}
0 & -i \Delta_\alpha \\ 
i \Delta^*_\alpha & 0
\end{array} \right ) L_0^\alpha\right ]\right ] \notag \\
&&- \frac{4Z_\omega}{\pi \gamma_c T} \sum_{\alpha} \Delta^*_\alpha \Delta_\alpha + S_{\rm int}^{s,t,c\vert_{n \neq 0}} .
\end{eqnarray}
The collection of interaction terms, Eqs.~\eqref{eq:Tstar:NLSM:Sintrho}-\eqref{eq:Tstar:NLSM:SintCooper}, (keeping only nonstatic parts in the Cooper channel) is abbreviated as $S_{\rm int}^{s,t,c\vert_{n \neq 0}} $.

As a next step, the space-dependent Bogoliubov rotation [analogous to \eqref{eq:Tstar:NLSM:BogoliubovTrafo}] is performed. For the derivation of interreplica terms in the GL functional, the fluctuations around the saddle point are parametrized by $q= W + \Lambda \sqrt{1 - W^2} \approx \Lambda + W - \Lambda W^2/2$ with $W$ Hermitian, $W \Lambda = - W \Lambda$ and $W  = t_{12} W^T t_{12}$ (same notation as in Refs.~\onlinecite{BurmistrovGornyiMirlin2012,BurmistrovGornyiMirlin2015}). This leads to 
\begin{eqnarray}
S &=& \frac{g}{32} \Tr \Big \lbrace (\nabla W)^2  +  \left [U_\Delta \nabla U_\Delta^\dagger,\Lambda\right ]^2 \notag \\
&& + 2 [W, \nabla W ]  U_\Delta \nabla U_\Delta^\dagger \notag \\
&&+ 2 \nabla W \left [U_\Delta \nabla U_\Delta^\dagger, \Lambda \right ] + 2 W \left [ \left [U_\Delta \nabla U_\Delta^\dagger,\Lambda \right ],U_\Delta \nabla U_\Delta^\dagger\right ] \notag \\
&&- \left [U_\Delta \nabla U_\Delta^\dagger, \Lambda \right ] \left [U_\Delta \nabla U_\Delta^\dagger, \Lambda W^2 \right ]  + \left [  U_\Delta \nabla U_\Delta^\dagger, W \right ]^2 \Big \rbrace \notag \\
&& - 2Z_\omega \Tr \left [ \hat \varepsilon \left (\Lambda - \frac{1}{2} \Lambda W^2 \right )\right ] - \frac{4Z_\omega}{\pi \gamma_c T} \sum_{\alpha} \vert \Delta_\alpha\vert^2 + S_{\rm int}^{{s,t,c\vert_{n \neq 0}}}\notag \\ \label{eq:GL:rotatedNLSM}
\end{eqnarray}
We first present the strategy for the calculation of $\langle \langle AA\rangle \rangle$ in the noninteracting case and consider interaction effects and other coefficients $\langle \langle AC\rangle \rangle$ and $\langle \langle CC\rangle \rangle$ afterwards.  

\subsection{Fluctuations in GL coefficient $A$: noninteracting case}

To obtain the coefficient $\langle \langle AA\rangle \rangle$, we can omit all $\nabla U_\Delta$ terms in Eq.~\eqref{eq:GL:rotatedNLSM} as well as the interaction terms. Then the action can be diagonalized with the parametrization 
\begin{equation}
W = \left (\begin{array}{cc}
0 & w \\ 
\bar w & 0
\end{array} \right ),
\end{equation}
where we expand ($\sigma_j= (\mathbf 1_\sigma, \vec \sigma)_j$)
\begin{subequations}
\begin{align}
w_{n_1,n_2}^{\alpha\beta} &= \frac{1}{2} \left (\begin{array}{cc}
\tilde d_{n_1n_2; j}^{\alpha \beta} & -i c_{n_1n_2; j}^{\alpha \beta}  \\ 
i \tilde c_{n_1n_2; j}^{\alpha \beta} & d_{n_1n_2; j}^{\alpha \beta} 
\end{array} \right )_{\tau} \sigma_j ,\\
\bar w_{n_2,n_1}^{\beta \alpha}  &= \frac{\lambda_j}{2} \left (\begin{array}{cc}
 d_{n_1n_2; j}^{\alpha \beta}& -i c_{n_1n_2; j}^{\alpha \beta} \\ 
i \tilde c_{n_1n_2; j}^{\alpha \beta} &\tilde d_{n_1n_2; j}^{\alpha \beta}
\end{array} \right )_{\tau} \sigma_j .
\end{align}
\end{subequations}
Here, the notation $\lambda_j$ means
 \begin{equation}
\lambda_j = \begin{cases}  -1, & \text{if } j \in \lbrace 1,2,3 \rbrace, \\ 1, & \text{else.} \end{cases}
\end{equation}
We use the convention that negative (zero or positive) Matsubara indices are denoted by even (respectively, odd) subscripts. The diagonal action is
\begin{eqnarray}
S_0[d,c] &=& \frac{g}{16} \int_{\v q} \sum_{\substack{n_1n_2\\ \alpha \beta \\0 \leq j \leq 3}} \lambda_j  \left [\mathcal D_\Delta^{-1}(\v q)\right ]_{n_1,n_2}^{\alpha \beta} \notag \\ &&  \left [\tilde d_{n_1n_2; j}^{\alpha \beta}d_{n_1n_2; j}^{\alpha \beta} + \tilde c_{n_1n_2; j}^{\alpha \beta}c_{n_1n_2; j}^{\alpha \beta}\right ], 
\label{eq:GL:deriv:diagoaction}
\end{eqnarray}
where 
\begin{equation}
\left [\mathcal D_\Delta^{-1}(\v q)\right ]_{n_1,n_2}^{\alpha \beta} = \v q^2 + \frac{\varepsilon_{n_1}^\alpha + \varepsilon_{n_2}^\beta}{D}\label{eq:GL:InversePropzero}
\end{equation}
and $D = g/16Z_\omega$. For convergence reasons, the complex diffuson fields have the property $\tilde d_j = \lambda_j d^*_j$ and analogously for cooperons. 

The integration of the modes yields a fluctuation determinant, or, equivalently, a fluctuation correction to the action
\begin{equation}
\delta S = 2 \bm{\beta} \sum_{\alpha, \beta} \sum_{n_1,n_2}  \Tr \ln \left [\mathcal D_\Delta^{-1}(\hat{\v q})\right ]_{n_1,n_2}^{\alpha \beta}. \label{eq:GL:deriv:Fluctaction}
\end{equation}
It is worth noticing that the action \eqref{eq:GL:deriv:diagoaction} is already diagonal in the multiindex $(\alpha , \beta , n_1, n_2)$. Thus the trace in Eq. \eqref{eq:GL:deriv:Fluctaction} does not include trace over replicas and/or Matsubara indices. However, the action \eqref{eq:GL:deriv:diagoaction} is not diagonal in momentum and coordinate space and the symbol ``$\Tr$'' refers to an operator trace in this space.

The parameter $\bm{\beta}$ is $\bm{\beta} = 4$ in the quaternionic case with spin-rotation invariance, with a singlet and  three triplet modes contributing. For broken spin-rotation invariance, when only the singlet mode contributes, we have $\bm{\beta} = 1$. 

We now expand the expression $ \varepsilon_n^\alpha = \sqrt{\epsilon_n^2 + \vert \Delta_\alpha (\hat{\v x}) \vert ^2}$ in small $\vert\Delta_\alpha\vert/T$. Recall that we are interested in \textit{inter}replica correction to the GL functional up to quartic order. It is thus sufficient to keep the quadratic order in the expansion of $\varepsilon_n^\alpha$ (the quartic order would lead to quartic \textit{intra}replica contributions with prefactor $N_R \rightarrow 0$)

\begin{equation}
 \varepsilon_n^\alpha \approx\vert  \epsilon_n \vert+ \frac{\vert \Delta^\alpha \vert^2}{2 \vert \epsilon _n \vert}.
\end{equation}

The expansion of $\varepsilon_n^\alpha$ leads to the approximate inverse propagator

\begin{equation}
\left [\mathcal D_\Delta^{-1}(\hat{\v q})\right ]_{n_1,n_2}^{\alpha \beta} = \mathcal D^{-1}(\hat{\v q}, \omega_{n_{12}})  + \frac{1}{D} \left [\frac{\vert \Delta^\alpha \vert^2}{2 \vert \epsilon _{n_1} \vert}+\frac{\vert \Delta^\beta \vert^2}{2 \vert \epsilon _{n_2} \vert}\right ]. 
\end{equation}
Here, $\omega_{n_{12}} = \epsilon_{n_1} - \epsilon_{n_2}$ and $\mathcal D$ denotes the propagator in a normal metal obtained from $\mathcal D_\Delta$ by setting $\Delta = 0$.

We expand the logarithm, Eq.~\eqref{eq:GL:deriv:Fluctaction}, to quartic order in $\Delta$ and keep only terms that survive the replica limit: 

\begin{eqnarray}
\delta S^{\rm GL} &=& - \bm{\beta} \sum_{\substack{\alpha, \beta\\ n_1, n_2}}  \int_{\v x, \v x'}  \frac{\vert \Delta ^\alpha \vert ^2_{\v x} \vert \Delta^\beta \vert ^2_{\v x'}}{2 D^2 \epsilon_{n_1} \vert \epsilon_{n_2} \vert} \notag \\
&& \times  \mathcal D( \v x- \v x', \omega_{n_{12}})\mathcal D( \v x'- \v x, \omega_{n_{12}})
\end{eqnarray}
As will be discussed below, the evaluation of the Matsubara sums leads to the correlation function for fluctuations of the GL coefficient $A$. 

\subsection{Fluctuation of GL coefficient C and interaction effects}

We now return to the derivation of the full, disordered GL functional, keeping terms up to quartic order in $\Delta/T$.
To this end, we consider Eq.~\eqref{eq:GL:rotatedNLSM}. It is apparent from the above derivation of the disorder correlations $\langle \langle A(\v x) A(\v x') \rangle \rangle$  that interaction terms $S_{\rm int}^{s,t,c\vert_{n \neq 0}}$ do not directly play a role for the derivation of the disordered GL functional. Indeed, interaction effects only appear in the propagators $\left [\mathcal D_\Delta^{-1}({\v q})\right ]_{n_1,n_2}^{\alpha \alpha}$. These terms cannot lead to interreplica terms, as is evident in view of Eq.~\eqref{eq:GL:deriv:Fluctaction}.
Further,  the first line of Eq.~\eqref{eq:GL:rotatedNLSM} produces the kinetic term of the normal state diffusion propagator and the standard (saddle point) kinetic term of the GL functional. The terms of the third line of Eq.~\eqref{eq:GL:rotatedNLSM} are unimportant, since they only affect the replica diagonal channel. Therefore, in addition to terms from $\Tr \hat \varepsilon \Lambda W^2$ we need to consider the following terms, of which the first two (last two) stem from line 2 (respectively, line 4) of Eq.~\eqref{eq:GL:rotatedNLSM}:
\begin{eqnarray}
\delta S[W,\Delta] &=& \frac{g}{32} \Tr \Big \lbrace  2 [ W, \nabla W]  U_\Delta \nabla U_\Delta^\dagger \notag \\
&&- \left [U_\Delta \nabla U_\Delta^\dagger, \Lambda \right ] \left [U_\Delta \nabla U_\Delta^\dagger, \Lambda W^2 \right ]  \notag \\
&& + \left [  U_\Delta \nabla U_\Delta^\dagger, W \right ]^2 \Big \rbrace \label{eq:GL:deriv:gradients}
\end{eqnarray}

As before, we expand to the second order in $\Delta/T$. This yields
\begin{equation}
U_\Delta \nabla U_\Delta^\dagger = -i \sum_{n\geq 0, \alpha} P_{\vert n \vert }^\alpha \left\lbrace \frac{ \Lambda_y}{2 \vert \epsilon_n \vert} \nabla \underline \Delta^\alpha + \vec{f}_n^\alpha \tau_z \right\rbrace. 
\end{equation}
Here we have introduced the following matrix in Nambu space:
\begin{equation}
\underline \Delta^\alpha  =  \left ( \begin{array}{cc}
0 & -i \Delta^\alpha \\ 
i \Delta^{\alpha,*} & 0
\end{array} \right ),
\end{equation}
and the vector field
\begin{equation}
\vec{f}_n^\alpha = \frac{i}{8 \epsilon_n^2} (\Delta^\alpha \nabla \Delta^{\alpha,*}-\Delta^{\alpha,*} \nabla \Delta^\alpha).
\end{equation}
We first consider the second and third term in the square bracket of Eq.~\eqref{eq:GL:deriv:gradients}. It is sufficient to keep terms up to linear order $\mathcal O(\Delta_\alpha^1)$ in $U_\Delta \nabla U_\Delta^\dagger $. 
We thus obtain from the second and third terms
\begin{eqnarray}
\delta S_{2,3} &=& \frac{g}{16}\sum_{\alpha; n \geq 0} \Tr \left [ P_{\vert n \vert}^{\alpha} \left ( \frac{-i \Lambda_y}{2 \vert \epsilon_n \vert} \nabla \underline \Delta^\alpha\right )^2 W^2 \right ] \notag \\
&+&  \frac{g}{16}\sum_{\substack{\alpha, \alpha'\\ n, n' \geq 0}} \Tr \left[ \left ( P_{\vert n \vert}^{\alpha} \frac{-i \Lambda_y}{2 \vert \epsilon_n \vert} \nabla \underline \Delta^\alpha\right ) W  \right.\notag \\
&& \left.\times \left ( P_{\vert {n'} \vert}^{\alpha'} \frac{-i \Lambda_y}{2 \vert \epsilon_{n'} \vert} \nabla \underline \Delta^{\alpha'}\right )  W \right ] .
\end{eqnarray}
The effect of these terms is twofold: The first term from $\delta S_{2,3}$ yields a shift of the propagator \eqref{eq:GL:InversePropzero} $\left [\mathcal D_\Delta^{-1}(\hat{\v q})\right ]_{n_1,n_2}^{\alpha \beta}  \rightarrow  \left [\tilde{\mathcal D}^{-1}(\hat{\v q})\right ]_{n_1,n_2}^{\alpha \beta} $ with
\begin{equation}
\left [\tilde{\mathcal D}^{-1}(\hat{\v q})\right ]_{n_1,n_2}^{\alpha \beta} \equiv  \left [\mathcal D_\Delta^{-1}(\hat{\v q})\right ]_{n_1,n_2}^{\alpha \beta}  - \frac{1}{4} \left ( \frac{\vert \nabla \Delta ^\alpha \vert^2}{\epsilon_{n_1}^2} + \frac{\vert \nabla \Delta ^\beta \vert^2}{\epsilon_{n_2}^2}\right ).  \label{eq:GL:FulDiagoProp}
\end{equation}
Second, the last terms from $\delta S_{2,3}$ produce couplings between cooperons and diffusons of opposite Matsubara frequencies:
\begin{eqnarray}
\delta S_{2,3}^b  &=& \frac{g}{16} \int_{\v x} \sum_{\substack{n_1, n_2\\ \alpha, \beta\\ 0 \leq  j \leq 3}} \frac{1}{4 \vert \epsilon_{n_1} \epsilon_{n_2} \vert} \notag \\
&\times& \left [\left (\begin{array}{c}
\tilde d_{n_1,n_2}^{\alpha \beta}\\ \tilde d _{-n_2-1, -n_1 - 1}^{\beta \alpha} \end{array}  \right )^T_j M^{(d)}_{\alpha,\beta} \left (\begin{array}{c}
 d_{n_1,n_2}^{\alpha \beta}\\  d _{-n_2-1, -n_1 - 1}^{\beta \alpha}
\end{array}  \right )_j \right.\notag \\
&&\left.+  \left (\begin{array}{c}
 \tilde c _{n_1,n_2}^{\alpha \beta}\\ c_{-n_2-1, -n_1 - 1}^{\beta \alpha}
\end{array}  \right )^T_j M^{(c)}_{\alpha,\beta}\left (\begin{array}{c}
 c _{n_1,n_2}^{\alpha \beta}\\ \tilde c_{-n_2-1, -n_1 - 1}^{\beta \alpha}
\end{array}  \right )_j \right ], \notag \\ \label{eq:GL:BouncingOfCondensateTerms}
\end{eqnarray}
with
\begin{subequations}
\begin{eqnarray}
M^{(d)}_{\alpha,\beta} &=& \left (\begin{array}{cc}
0 & \nabla \Delta^{*,\alpha} \nabla \Delta^\beta \\ 
\nabla \Delta^{\alpha} \nabla \Delta^{*,\beta} & 0 
\end{array} \right ) , \\
M^{(c)}_{\alpha,\beta} &=&\left (\begin{array}{cc}
0 & \nabla \Delta^{\alpha} \nabla \Delta^\beta \\ 
 \nabla \Delta^{*,\alpha} \nabla \Delta^{*,\beta} & 0
\end{array} \right ) .
\end{eqnarray}
\end{subequations}

We now consider the first term (with the single commutator) in the square bracket of Eq.~\eqref{eq:GL:deriv:gradients}.
Since $W^2$ and $W \nabla W$ are block diagonal in Matsubara space, $\Tr \Lambda_y W^2 = 0 = \Tr \Lambda_y W \nabla W$. Thus only terms stemming from the term proportional to $\vec f_n^\alpha$ in $U_\Delta \nabla U_\Delta^\dagger$ survive the trace operation. 
The first term thus yields the contribution
\begin{eqnarray}
&&\delta S_1 =  \frac{g}{32} \sum_{j = 0 , \dots, 3}\lambda_j  \notag \\
&&\int_{\v p, \v q}\Big \lbrace \tilde d^{\alpha \beta}_{n_1n_2;j} \left [ \vec f_{n_1}^\alpha ( \v q) - \vec f_{n_2}^\beta ( \v q) \right ] (2  \v p + \v q ) d^{\alpha \beta}_{n_1n_2;j}  \notag \\
&& + \tilde c^{\alpha \beta}_{n_1n_2;j} \left [ \vec f_{n_1}^\alpha ( \v q) + \vec f_{n_2}^\beta ( \v q) \right ] (2  \v p + \v q ) c^{\alpha \beta}_{n_1n_2;j} \Big \rbrace \notag \\ 
\end{eqnarray}
In the replica limit we will be only interested in contributions from the crossed terms which cancel out between cooperons and diffusons in view of the different sign of contributions. We therefore do not need to consider $\delta S_1$ any longer.

The fluctuation determinant leads thus to a correction of GL action of the form
\begin{eqnarray}
\delta S^{\mathrm{GL}} &=& \frac{1}{2} \sum_{\alpha, \beta, n_1, n_2;j} \notag \\ 
&& \Tr \ln \left [\left [D^{-1}(\hat{\v q})\right ]_{n_1,n_2}^{\alpha \beta} + \frac{(-\lambda_j)}{2 \epsilon_{n_1} \vert \epsilon_{n_2} \vert}M^{(d)}_{\alpha,\beta} \right ] \notag \\
&+& \frac{1}{2} \sum_{\alpha, \beta, n_1, n_2;j}  \notag \\ 
&&\Tr \ln \left [\left [D^{-1}(\hat{\v q})\right ]_{n_1,n_2}^{\alpha \beta} + \frac{(-\lambda_j)}{2 \epsilon_{n_1} \vert \epsilon_{n_2} \vert} M^{(c)}_{\alpha,\beta}\right ] . \nonumber \\
&& 
\end{eqnarray}
Here, the first two lines stem from diffusons and the third and fourth line from cooperons. The symbol $\Tr$ denotes trace in momentum/coordinate space and in the space of the $2\times 2$ matrices introduced in Eq. ~\eqref{eq:GL:BouncingOfCondensateTerms}. It does not include the trace over replica and/or Matsubara indices. The expansion of the trace to fourth order in $\Delta$ leads to
\begin{eqnarray}
\delta S^{\rm GL}_{\rm dis} &=& -\sum_{\substack{\alpha, \beta\\ n_1, n_2}}  \int_{\v x, \v x'} \mathcal D( \v x- \v x', \omega_{n_{12}})\mathcal D( \v x'- \v x, \omega_{n_{12}}) \notag \\
&& \times  \bm{\beta} \left\lbrace \frac{\vert \Delta ^\alpha \vert ^2_{\v x} \vert \Delta^\beta \vert ^2_{\v x'}}{2 D^2 \epsilon_{n_1} \vert \epsilon_{n_2} \vert}- \frac{\vert \nabla \Delta ^\alpha \vert ^2_{\v x} \vert \Delta^\beta \vert ^2_{\v x'}}{2 D \epsilon_{n_1} \vert \epsilon_{n_2} \vert^2} \right.\notag \\
&& + \frac{1}{8 \epsilon_{n_1}^2\epsilon_{n_2}^2} \Big [ \vert \nabla \Delta ^\alpha \vert ^2_{\v x} \vert \nabla \Delta^\beta \vert ^2_{\v x'} \notag \\
&&\phantom{+ \frac{1}{8 \epsilon_{n_1}^2\epsilon_{n_2}^2} } + (\nabla \Delta^*_\alpha  \nabla \Delta_\beta )_{\v x}  (\nabla \Delta_\alpha  \nabla \Delta^*_\beta )_{\v x'} \notag \\
&&\left.\phantom{+ \frac{1}{8 \epsilon_{n_1}^2\epsilon_{n_2}^2} }  + (\nabla \Delta_\alpha  \nabla \Delta_\beta )_{\v x}  (\nabla \Delta_\alpha^*  \nabla \Delta^*_\beta )_{\v x'} \Big] \right \rbrace.\nonumber \\
&& 
\end{eqnarray}
This expression is the origin of Eq.~\eqref{eq:GL:CollectionInterreplica} of the main text.

\subsection{Discussion of disordered GL functional}

All in all, our derivation yields the following interreplica corrections to the GL functional to the leading order in the expansion $\vert \Delta \vert/T$:
\begin{eqnarray}
\delta S^{\rm GL}_{\rm dis} &=& - \frac{1}{2T^2}  \sum_{\alpha \beta} \int_{\v x, \v x'} \Big \lbrace \langle \langle A(\v x) A(\v x') \rangle \rangle  \vert \Delta_\alpha \vert ^2_{\v x} \vert \Delta_\beta \vert ^2_{\v x'} \notag \\
&&  + 2\langle \langle A(\v x) C(\v x') \rangle \rangle  \vert \partial_i \Delta_\alpha \vert ^2_{\v x} \vert \Delta_\beta \vert ^2_{\v x'} \notag \\
&& + \frac{\langle \langle C(\v x) C(\v x') \rangle \rangle}{2} \Big [ \vert \partial_i \Delta_\alpha \vert ^2_{\v x} \vert \partial_i \Delta_\beta \vert ^2_{\v x'} \notag \\
&& \phantom{ \frac{\langle \langle C(\v x) C(\v x') \rangle \rangle}{2} } + (\nabla \Delta^*_\alpha  \nabla \Delta_\beta )_{\v x}  (\nabla \Delta_\alpha  \nabla \Delta^*_\beta )_{\v x'} \notag \\
&& \phantom{\frac{\langle \langle C(\v x) C(\v x') \rangle \rangle}{2} } + (\nabla \Delta_\alpha  \nabla \Delta_\beta )_{\v x}  (\nabla \Delta_\alpha^*  \nabla \Delta^*_\beta )_{\v x'}  \Big ]
\Big \rbrace . \notag \\
\label{eq:GL:App:CollectionInterreplica}
\end{eqnarray}

We will be interested in the limit when the field $\Delta$ is smooth on the length scale of $L_T$.
In Eq.~\eqref{eq:GL:CollectionInterreplica} of the main text we presented only the delta correlated part of fluctuating GL parameters. Here we will go beyond this approximation. 
In general, the Gaussian statistics of GL parameters is dictated by mean values given in Eqs.~\eqref{eq:GL:parametersGL} and fluctuations
\begin{align}
&\left\langle\left \langle \left (\begin{array}{c}
A( \v x) \\ 
\frac{2 \pi T}{D} C( \v x)
\end{array} \right ) ^T \left (\begin{array}{c}
A( \v x') \\ 
\frac{2 \pi T}{D} C( \v x')
\end{array} \right ) 
 \right \rangle \right \rangle \notag \\
 &= \frac{\bm{\beta}}{D (2\pi)^3 T} \left (\begin{array}{cc}
\gamma_{AA}( \v x- \v x') & - \gamma_{AC}( \v x- \v x') \\ 
- \gamma_{AC}( \v x- \v x')&\gamma_{CC}( \v x- \v x')
\end{array} \right ) \label{eq:GL:app:MatrixCorrelfunction}
\end{align}
It is convenient to present the correlation functions entering Eq.~\eqref{eq:GL:app:MatrixCorrelfunction}  in Fourier space:
\begin{subequations}
\begin{eqnarray}
\gamma_{AA}( \v q)&=& \int_{0}^1\frac{ du}{4\pi} \sum_{n_1,n_2} \frac{1}{D^{-1} \omega_{n_{12}} + \v q^2 (u-u^2)} \frac{(2 \pi T)^3}{D \epsilon_{n_1} \vert \epsilon_{n_2}\vert} \notag \\
&\approx & \frac{1}{4\pi } \left (S_1^{(a)}  - \frac{1}{6} S_1^{(b)} \frac{D\v q^2}{2\pi T} + \frac{1}{30} S_1^{(c)}\left [ \frac{D\v q^2}{2\pi T}\right ]^2   \right ),\notag \\ &&\label{eq:GL:FluctuationTC:FT}\\
 \gamma_{AC}( \v q)&=& \int_{0}^1\frac{ du}{4\pi} \sum_{n_1,n_2} \frac{1}{D^{-1} \omega_{n_{12}} + \v q^2 (u-u^2)} \frac{(2 \pi T)^4}{2D \epsilon_{n_1} \epsilon_{n_2}^2} \notag \\ 
 &\approx & \frac{1}{8\pi } \left (S_2^{(a)} - \frac{1}{6} S_2^{(b)} \frac{D \v q^2}{2\pi T} \right ), \label{eq:GL:MixedFluct:FT} \\
\gamma_{CC}( \v q)&=& \int_{0}^1\frac{ du}{4\pi} \sum_{n_1,n_2} \frac{1}{D^{-1} \omega_{n_{12}} + \v q^2 (u-u^2)} \frac{ 2 (2 \pi T)^5}{4 D \epsilon_{n_1}^2 \epsilon_{n_2}^2}\notag \\
&\approx &  \frac{1}{8 \pi }S_3 . \label{eq:GL:ConductivityFluct:FT} 
\end{eqnarray}
\label{eq:GLApp:CollectionCorrelators}
\end{subequations}
In the approximate evaluation, we only kept terms yielding terms $\partial^n(\Delta/T)^m$ with $n, \leq m \leq 4$ in the GL functional. We will use the notation $\gamma_{AA}$ for $\gamma_{AA}( \v q = 0)$ and analogously for $\gamma_{AC}$, $\gamma_{CC}$. The sums entering these expressions are
\begin{subequations}
\begin{eqnarray}
S_1^{(a)} &= & \sum_{n_1,n_3} \frac{1}{(n_1 + n_3 + 1)(n_1 + 1/2)(n_3+1/2)} = 7 \zeta(3)\notag \\
&& \\
S_1^{(b)} &= & \sum_{n_1,n_3} \frac{1}{(n_1 + n_3 + 1)^2(n_1 + 1/2)(n_3+1/2)} \approx 5.2,\notag \\
&& \\
S_1^{(c)} &= & \sum_{n_1,n_3} \frac{1}{(n_1 + n_3 + 1)^3(n_1 + 1/2)(n_3+1/2)} \approx 4.5,\notag \\
&& \\
S_2^{(a)} &= & \sum_{n_1,n_3} \frac{1}{(n_1 + n_3 + 1)(n_1 + 1/2)(n_3+1/2)^2} = \frac{\pi^4}{8}\notag \\
&& \\
S_2^{(b)} &= & \sum_{n_1,n_3} \frac{1}{(n_1 + n_3 + 1)^2(n_1 + 1/2)(n_3+1/2)^2}= S_3/2\notag \\
&&\\
S_3 &= & \sum_{n_1,n_3} \frac{1}{(n_1 + n_3 + 1)(n_1 + 1/2)^2(n_3+1/2)^2}\notag \\
&=& \ 7 \pi^2 \zeta(3) - 62 \zeta(5) 
\end{eqnarray}
\end{subequations}

The eigenvalues of the matrix of $\hat{\gamma}$ are positive in the limit $D \v q ^2/T \ll 1$. At $\v q =0$ they are approximately $15/4\pi$ and $2.8/4\pi$. In the main text, we further used
\begin{align}
&(\partial_i \Delta^*_\alpha  \partial_i \Delta_\beta ) (\partial_j \Delta_\alpha  \partial_j \Delta^*_\beta )  \notag \\
&+ (\partial_i \Delta_\alpha  \partial_i \Delta_\beta )  (\partial_j \Delta^*_\alpha  \partial_j \Delta^*_\beta ) =\notag \\
&(\partial_i \Delta^*_\alpha  \partial_i \Delta_\alpha ) (\partial_j \Delta_\beta  \partial_j \Delta^*_\beta )  \notag \\
&+(\partial_i \Delta^*_\alpha [s_x]_{ii'}  \partial_{i'} \Delta_\alpha ) (\partial_j \Delta_\beta [s_x]_{jj'}   \partial_{j'} \Delta^*_\beta ) \notag \\
&+(\partial_i \Delta^*_\alpha [s_z]_{ii'}  \partial_{i'} \Delta_\alpha ) (\partial_j \Delta_\beta [s_z]_{jj'}   \partial_{j'} \Delta^*_\beta ).
\end{align}

\section{Derivation of the thermal conductivity}
\label{app:Kappa}

In this section, we present the derivation of the quasiparticle contribution to thermal conductance in the disordered superconductor. Upon taking $\Delta \rightarrow 0$, this discussion also treats the normal conducting case.

The thermal conductance is calculated on the level of the Kubo-Formula introduced in Ref.~\onlinecite{Luttinger-ThermalTransport} for both normal metals and superconductors. In the conductivity bubble, the electrical charge in each current vertex is replaced by the half sum of adjacent fermionic Matsubara frequencies. Within the superconducting, interacting NL$\sigma$M approach (see Sec.~\ref{sec:SCNLSM}), this amounts to introducing an energy dependent vector potential as a source field\cite{footnoteHeatVertex}
\begin{equation}
S_\sigma = \frac{g}{32} \int_{\v x} \tr [(D_i Q)^2] - 2Z_\omega \int_{\v x} \tr[ (\hat \epsilon  + \sum_\alpha \Delta \tau_y L_{\alpha,0}) Q].
\end{equation}
The covariant derivative is 
\begin{equation}
D_i = \partial_i +i \left [A_{i,n}^{\alpha} \left (\begin{array}{cc}
-[\mathcal I_n^{\alpha}]^T &  \\ 
 & \mathcal I_n^{\alpha}
\end{array} \right )_\tau,\bullet\right ].
\end{equation}
For the calculation of electrical conductivity, one replaces $\mathcal I_n^\alpha \rightarrow e I_n^\alpha$, where $e$ is the electrical charge. Instead, for the calculation of thermal conductivity, we introduce
\begin{equation}
\left (\mathcal I_{n_0}^{\alpha_0}\right )^{\alpha \beta}_{nm} = \delta^{\alpha_0\alpha}  \delta^{\alpha_0\beta} \delta_{n-m,n_0} i \frac{\epsilon_n+\epsilon_m}{2}.
\end{equation}

To obtain the thermal conductance including all quantum corrections stemming from scales $L\in(l,L_{T_c})$ we perform RG up to the infrared cut-off scale and then evaluate the gradient term of the NL$\sigma$M in the saddle point approximation $Q = \bar \Lambda$. We use
\begin{eqnarray}
&& \tr\left [\left (\begin{array}{cc}
-[\mathcal I_n^{\alpha}]^T &  \\ 
 & \mathcal I_n^{\alpha}
\end{array} \right )_\tau \bar \Lambda\left (\begin{array}{cc}
-[\mathcal I_{-n}^{\alpha}]^T &  \\ 
 & \mathcal I_{-n}^{\alpha}
\end{array} \right )_\tau \bar \Lambda \right ] \notag \\
&&= 2\sum_{k,l} \left \lbrace [\mathcal I_n^\alpha]^T_{kl} \frac{\epsilon_l}{\sqrt{\epsilon_l^2 + \Delta^2}}[\mathcal I_{-n}^\alpha]^T_{lk} \frac{\epsilon_k}{\sqrt{\epsilon_k^2 + \Delta^2}} \right.\notag \\
&&+  [\mathcal I_n^\alpha]_{kl} \frac{\epsilon_l}{\sqrt{\epsilon_l^2 + \Delta^2}}[\mathcal I_{-n}^\alpha]_{lk} \frac{\epsilon_k}{\sqrt{\epsilon_k^2 + \Delta^2}} \notag \\
 &&-[\mathcal I_n^\alpha]^T_{kl} \frac{\Delta}{\sqrt{\epsilon_l^2 + \Delta^2}}[\mathcal I_{-n}^\alpha]_{-(l+1),-(k+1)} \frac{\Delta}{\sqrt{\epsilon_k^2 + \Delta^2}} \notag \\
 &&\left.- [\mathcal I_n^\alpha]_{kl} \frac{\Delta}{\sqrt{\epsilon_l^2 + \Delta^2}}[\mathcal I_{-n}^\alpha]^T_{-(l+1),-(k+1)} \frac{\Delta}{\sqrt{\epsilon_k^2 + \Delta^2}}\right \rbrace
\end{eqnarray} 
to determine the effective action for the source fields of the heat current
\begin{eqnarray}
&&S_{\rm heat}[A_{i,n}^\alpha] = - \frac{g}{4} \sum_{\alpha,n} A_{i,n}^\alpha A_{i,-n}^\alpha\nonumber\\
&&\sum_{k} \left [ \frac{\epsilon_k \epsilon_{k+n} + \Delta^2}{\sqrt{\epsilon_k^2 + \Delta^2}\sqrt{\epsilon_{k+n}^2 + \Delta^2}}-1\right ]\left [i \frac{\epsilon_k +\epsilon_{k+n}}{2}\right ]^2.
\end{eqnarray}
For comparison, we present the same formula for the source fields of the electric current
\begin{eqnarray}
S_{\rm el}[A_{i,n}^\alpha] = - \frac{g}{4} \sum_{\alpha,n} e^2A_{i,n}^\alpha A_{i,-n}^\alpha\nonumber\\
\sum_{k} \left [ \frac{\epsilon_k \epsilon_{k+n} - \Delta^2}{\sqrt{\epsilon_k^2 + \Delta^2}\sqrt{\epsilon_{k+n}^2 + \Delta^2}}-1\right ],
\end{eqnarray}
which can be evaluated to obtain the Mattis-Bardeen ac-conductivity formula in the superconductor including all quantum corrections stemming from scales $L \in \{l, \min (L_\omega,L_{T_c})\}$. When evaluating this expression in the limit $\Delta \rightarrow 0$ we obtain 
\begin{equation}
S_{\rm el} = \sum_{\alpha,n}\frac{g e^2}{2\pi} \frac{\vert \omega_n \vert}{2T} A_{i,n}^\alpha A_{i,-n}^\alpha
\end{equation}
i.e. the normal state dc conductivity of $\sigma = g e^2/2\pi = g e^2/h$.

We proceed with the evaluation of the thermal conductivity $\kappa$ entering the effective action of source fields as
\begin{equation}
S_{\rm heat} = \sum_{\alpha,n} \frac{\kappa}{T} \frac{ \vert \omega_n \vert}{2T} A_{i,n}^\alpha A_{i,-n}^\alpha,
\end{equation}
and thus being for the superconductor (for simplicity we concentrate on $\omega_n >0$) 
\begin{eqnarray}
&&\kappa=\frac{g }{2 \omega_n} \sum_{k} \left [ \frac{\epsilon_k \epsilon_{k+n} + \Delta^2}{\sqrt{\epsilon_k^2 + \Delta^2}\sqrt{\epsilon_{k+n}^2 + \Delta^2}}-1\right ]\left [\frac{\epsilon_k +\epsilon_{k+n}}{2}\right ]^2  \nonumber\\
&&\stackrel{\Delta \rightarrow 0}{=}  \frac{-g}{\omega_n} (\pi T)^2 \underbrace{\sum_{k =0}^{n-1} (2k +1 - n)^2}_{ = (-n+n^3)/3} = \frac{g}{2\pi} \frac{\pi^2}{3} T \label{eq:thermalconductance}
\end{eqnarray}
where we analytically continued $\omega_n \rightarrow \omega+ i 0$. We thus recovered the Wiedemann-Franz law in the normal state (but including renormalization\cite{SchwieteFinkelstein2014}).

Now let us consider the superconducting state. First, we note that the sum in Eq.~\eqref{eq:thermalconductance} is formally UV divergent. This is a well known property of the Kubo expression for the thermal conductance in superconductors. This divergence is cancelled by terms stemming from the time derivative of time ordering Heaviside functions (see e.g. discussion in Ref.~\onlinecite{Mineev}) which is equivalent to disregarding the large $\epsilon_k$ contribution in the contour integration.

We define
\begin{equation}
f(i \epsilon_k, i \epsilon_{k+n}) =\left [ \frac{\epsilon_k \epsilon_{k+n} + \Delta^2}{\sqrt{\epsilon_k^2 + \Delta^2}\sqrt{\epsilon_{k+n}^2 + \Delta^2}}-1\right ]\left [\frac{\epsilon_k +\epsilon_{k+n}}{2}\right ]^2
\end{equation}
and evaluate the sum $S(\omega)$ entering $\kappa = \frac{g}{-i2 \omega} S(\omega+i0) \vert_{\omega \rightarrow 0}$
\begin{eqnarray}
&&S(i \omega_n) = \sum_k f(i \epsilon_k, i \epsilon_{k+n})
= \int_\Delta^\infty d \epsilon\frac{\tanh(\epsilon/2T)}{2\pi i T } \nonumber\\
&&\Big [f(\epsilon + i 0,\epsilon + i \omega_n) - f(\epsilon - i 0,\epsilon + i \omega_n) \notag \\
&&+f(\epsilon + i 0,\epsilon - i \omega_n) - f(\epsilon - i 0,\epsilon - i \omega_n)  \Big ].
\end{eqnarray}
It follows that
\begin{eqnarray}
&&S(\omega + i0) = \int_\Delta^\infty d \epsilon\frac{\tanh([\epsilon+\omega]/2T)-\tanh(\epsilon/2T)}{\pi i T } \nonumber\\
&&\left  (\epsilon + \frac{\omega}{2}\right )^2 \left [ \frac{\epsilon (\epsilon + \omega) - \Delta^2}{\sqrt{\epsilon^2 - \Delta^2}\sqrt{(\epsilon+\omega)^2 - \Delta^2}}\right ] \notag \\ 
&&-i \int_{\Delta-\omega}^\Delta d \epsilon\frac{\tanh([\epsilon+\omega]/2T)}{\pi i T } \nonumber\\
&&\left  (\epsilon + \frac{\omega}{2}\right )^2 \left [ \frac{\epsilon (\epsilon + \omega) - \Delta^2}{\sqrt{ \Delta^2-\epsilon^2}\sqrt{(\epsilon+\omega)^2 - \Delta^2}}\right ]
\end{eqnarray}
Expanding in $\omega$ we obtain Eq. (\ref{eq:ThermalConductivity}) from the main text. On the Drude level for electrons with quadratic dispersion we have $g/2\pi = n \tau/m$, and thus our result reduces to the formula of Ambegaokar and Griffin.~\cite{AmbegaokarGriffin}



\begin{thebibliography}{100}
\bibitem{Footnote-Gi}
There exist different definitions for $Gi$ in the literature. Here, it is defined via the temperature $T_{\rm BCS}(1+ Gi)$ at which the fluctuation correction to heat capacity is comparable to the mean field jump in the same observable. The ratio of $Gi$ in this convention as compared to the definition by means of the scale where the Aslamazov-Larkin contribution reaches the value of $g_D$ is given by the parameter $A_v$ in Eq.~\eqref{eq:NelsonHalperin}.
\bibitem{Footnote-Q}
In the chosen basis, matrices $Q$ are symplectic (class AI) or orthogonal (class AII) up to a unitary rotation.
\bibitem{Footnote-Matsubara}
Since the fermionic Matsubara frequencies are defined as $\epsilon_n = \pi T ( 2n + 1)$ the constraint $n+n'+1 = m$ corresponds to $\epsilon_n + \epsilon_{n'} = \omega_m$.
\bibitem{FootnoteKbare}
In the limit when thermal fluctuations dominate, the mean stiffness is given by $K = \pi \tilde C \langle \vert \Delta \vert^2 \rangle/T$ with $\langle \dots \rangle $ denoting the average with respect to $S[ \vert \Delta \vert ] =  \int_{\v x} \lbrace \tilde A \vert \Delta \vert^2 + \tilde B \vert \Delta \vert^4/2 + \tilde C (\nabla \vert \Delta \vert)^2 \rbrace/T$. Rescaling of the amplitude field by $\sqrt{-\tilde B/ \tilde A}$ and of spatial coordinates by $\sqrt{\tilde A^2/\tilde BT}$ indicates that $K(\vert \epsilon \vert, Gi_X) = K(Gi_X/\vert \epsilon \vert)$.
\bibitem{Footnote-LBKT}
On the normal conducting side of the transition, we define the BKT correlation length as the scale at which Eqs.~\eqref{eq:standardBKTRG} predict $\bar K=0$.
\bibitem{footnoteHeatVertex}
{Calculating the thermal conductivity by means of ``energy dependent vector potentials'' is  justified in the non-interacting case and for the saddle-point evaluation of $\kappa$ in interacting systems (the case considered here).}
\end{thebibliography}
\end{document}